\documentclass[twocolumn,floats,floatfix,amssymb,prd,superscriptaddress,nofootinbib,preprintnumbers,notitlepage]{revtex4-1}

\usepackage{subcaption}
\usepackage{ragged2e}
\DeclareCaptionJustification{justified}{\justifying}
\makeatletter
\newcommand{\subsetsim}{\mathrel{\mathpalette\subset@sim\relax}}
\newcommand{\subset@sim}[2]{%
  \vtop{\offinterlineskip\m@th
    \ialign{\hfil##\cr
      $#1\subset$\cr\noalign{\kern0.5pt}\scalebox{0.9}{$#1\sim$}\cr
    }%
  }%
}
\makeatother

\usepackage{amssymb,amsmath,verbatim,mathtools,needspace,enumitem,etoolbox,graphicx,physics,microtype,afterpage,bm}
\usepackage[dvipsnames, usenames]{xcolor}
\definecolor{linkcolor}{rgb}{0.0,0.3,0.5}
\usepackage{booktabs}
\usepackage[unicode, colorlinks=true, linkcolor=linkcolor, citecolor=linkcolor, filecolor=linkcolor,urlcolor=linkcolor, pdfusetitle]{hyperref}
\usepackage[all]{hypcap}
\usepackage[T1]{fontenc}
\usepackage[utf8]{inputenc}
\usepackage{tabularx}
\usepackage{float}
\usepackage{cleveref}
\usepackage{comment}
\usepackage{multirow}
\usepackage{pifont}
\usepackage{lmodern}

\allowdisplaybreaks
\usepackage{tikz}
\usepackage{framed}
\usepackage{hyperref}
\hypersetup{colorlinks, citecolor=bluscuro, linkcolor=black, urlcolor=bluscuro}
\definecolor{bluscuro}{rgb}{0.15, 0.2, .85}
\definecolor{ForestGreen}{rgb}{0.13, 0.55, 0.13}

\newcommand{\beq}{\begin{equation}}
\newcommand{\eeq}{\end{equation}}
\newcommand{\barr}{\begin{eqnarray}}
\newcommand{\earr}{\end{eqnarray}}

\renewcommand{\d}{{\rm d}}

\newcommand{\be}{\begin{equation}}
\newcommand{\ee}{\end{equation}}
\newcommand{\tn}{\textnormal}

\newcommand{\cern}{
CERN, Theoretical Physics Department,
Esplanade des Particules 1, Geneva 1211, Switzerland}

\begin{document}

\title{
Gravitational Wave Memory of Primordial Black Hole Mergers
}

\author{Silvia Gasparotto}
\email{sgasparotto@ifae.es}
\affiliation{Institut de F\'isica d’Altes Energies (IFAE), The Barcelona Institute of Science and Technology, Campus UAB, 08193 Bellaterra (Barcelona), Spain}
\affiliation{Grup de F\'{i}sica Te\`{o}rica, Departament de F\'{i}sica, Universitat Aut\`{o}noma de Barcelona, 08193 Bellaterra (Barcelona), Spain}{}

\author{Gabriele Franciolini}
\email{gabriele.franciolini@cern.ch}
\affiliation{\cern} 

\author{Valerie Domcke}
\email{valerie.domcke@cern.ch}
\affiliation{\cern}

\begin{abstract}
The gravitational wave signal of binary compact objects has two main contributions at frequencies below the characteristic merger frequency: the gravitational wave signal associated with the early inspiral stage of the binary and the non-linear gravitational wave memory. We compare the sensitivity of upcoming gravitational wave detectors to these two contributions, with a particular interest in events with a merger phase at frequencies higher than the detector's peak sensitivity. We demonstrate that for light primordial black holes, current and upcoming detectors are more sensitive to the inspiral signal. Our analysis incorporates the evolution history of primordial black hole binaries, key to accurately estimating the relevant event rates. We also discuss the waveform templates of the memory signal at ground- and space-based interferometers, and the implications for a matched filtering search. This allows us to compare the sensitivity of high-frequency gravitational wave detectors, sensitive to the merger phase, with the sensitivity of existing interferometers.
\end{abstract}

\maketitle

\preprint{CERN-TH-2025-051} 


\section{Introduction}
\label{sec:introduction}

The first detection of gravitational waves (GWs) by the LIGO-Virgo-KAGRA (LVK) network marked the dawn of a revolutionary era in astronomy~\cite{LIGOScientific:2016aoc}. With over 90 GW events now identified by the LVK collaboration~\cite{KAGRA:2021vkt} and the recent indication of a stochastic GW background at pulsar timing arrays~\cite{Xu:2023wog,EPTA:2023fyk,NANOGrav:2023gor,Reardon:2023gzh,Miles:2024rjc}, our understanding of the universe’s most dynamic and energetic events has expanded dramatically. These discoveries have opened a new window into the cosmos, shedding light on the population and properties of their sources~\cite{EPTA:2023xxk}.

A major challenge in this field is to unravel the origin of binaries of compact objects. Such systems may arise either from astrophysical processes in the late universe or from primordial density fluctuations in the early universe, potentially leading to the formation of Primordial Black Holes (PBHs)~\cite{Zeldovich:1967lct,Hawking:1971ei,Carr:1974nx, Carr:1975qj} (see e.g.~\cite{Byrnes:2025tji} for recent reviews). PBHs, remnants of the early universe, offer an exciting opportunity to probe fundamental physics. They may also constitute a significant fraction of the universe’s Dark Matter (DM) content (see~\cite{Carr:2020gox} for a review of current constraints) and lead to detectable GW merger events \cite{Bird:2016dcv,Clesse:2016vqa,Sasaki:2016jop,Eroshenko:2016hmn, Wang:2016ana, Ali-Haimoud:2017rtz,Chen:2018czv,Raidal:2018bbj,Franciolini:2021tla,Franciolini:2021nvv,Afroz:2024fzp}.

A detection of a sub-solar black hole (BH) would provide direct evidence for a cosmological origin and/or new physics, as such objects cannot be produced by known astrophysical mechanisms (see e.g. \cite{Crescimbeni:2024qrq}).
PBHs in the asteroid-mass range, $m_{\rm PBH} \in [10^{-16}, 10^{-10}] \, M_\odot$, are especially compelling as they could account for the entirety of DM \cite{Carr:2020gox}. However, detecting the gravitational wave signal from light primordial black hole binaries is challenging: the signal strength scales with the mass of the binary, and the maximal strain is achieved at frequencies above the reach of the LVK network. This makes them a prime target for high-frequency GW (HFGW) detectors, which however currently do not have sufficient sensitivity to constrain this parameter space~\cite{Franciolini:2022htd,Aggarwal:2025noe}.

In this situation, we focus on the low-frequency tail of the GW signal of light PBH binaries, which may be probed by the LVK network or upcoming GW detectors such as the Einstein Telescope (ET) \cite{Franciolini:2023opt,Abac:2025saz} and LISA \cite{LISACosmologyWorkingGroup:2023njw,LISA:2024hlh}. PBH binaries generate a GW signal at frequencies far below the peak merger signal at two distinct phases of their evolution. First, during their early inspiral, they emit quasi-monochromatic GWs. 
Second, another low-frequency GW signature, known as the \textit{GW memory}~\cite{Christodoulou:1991cr,Blanchet:1992br,Thorne:1992sdb} (see \cite{Mitman:2024uss} for a recent review), is generated by all GW sources and reaches its maximum strength during the merger phase for a binary coalescence.

This memory effect, a second-order GW phenomenon, is associated with the passage of the primary GW energy flux. It leaves a permanent offset in the GW strain, representing a transition between two distinct flat spacetime states before and after the event. The memory signal appears as a step-like function at the merger time, with its Fourier transform displaying a $\sim f^{-1}$ dependence at low frequencies. The significant low-frequency contribution of this signal offers a compelling pathway to study high-frequency mergers indirectly using low-frequency GW detectors. In particular,~\cite{McNeill:2017uvq} suggests that searches for ``orphan memory'' in LVK, that is, GW memory associated with out-of-band primary signals, could surpass direct searches in HFGW detectors for burst-like high-frequency events. Their analysis is based on sine-gaussian signals which, contrary to PBH mergers, do not have a low-frequency component besides the memory signal.

In this paper, we critically re-evaluate these claims in the context of PBH populations. Specifically, we compare the ability of low-frequency detectors to detect both the inspiral and memory signals from PBH mergers with the performance of proposed
HFGW detectors focused on the final merger stages. We incorporate the full evolutionary trajectory of PBH populations to evaluate the expected number of binaries emitting at a given frequency, including also those very far away from their merger, becoming nearly monochromatic sources. We find this to be crucial for an accurate prediction of the sensitivity to the inspiral signal.

Searches for both the primary and memory signal benefit 
from the implementation of a matched filtering search, i.e.\ the use of a signal template. For the primary signal this is a well-established technique~\cite{maggiore2008gravitational}. For the memory signal, the properties of the signal templates depend crucially on the detector response function, see Ref.~\cite{Ghosh:2023rbe} for LISA and Ref.~\cite{Ebersold:2020zah} for an approximation of memory templates for LIGO. Remarkably, the resulting waveform can be to good approximation characterized by very few parameters. We revisit and refine these derivations, including the full instrument response function and discussing the implications for a matched filtering search.

The structure of the paper is as follows: In \Cref{sec:GWsig}, we review the GW memory effect associated with BH mergers, modeling the GW signal and comparing it with the inspiral GW emission.  Section~\ref{sec:PBHmergerrate} discusses the PBH merger rates used in our analysis. Section~\ref{sec:res} details our methodology and results before we conclude in \Cref{sec:conclusions}. The appendices contain: \ref{app:memtheory}) a derivation of the GW memory signal;
\ref{app:response}) a derivation of the memory signal templates at interferometers; \ref{app:sens_curves}) the sensitivity curves of the detectors considered in this paper. 
In what follows we set $c=G=1$.


\section{GWs from BH binaries}\label{sec:GWsig}

In this section, we discuss the GW signatures of BH mergers, with a particular focus on the GW memory.

\subsection{GW memory signal and modeling}\label{sec:GWmem}

The gravitational wave memory effect refers to the non-oscillatory component of a GW signal, characterized by the permanent offset in the GW strain between the start and end of the wave's passage. This phenomenon can be categorized into two main types.

The first, known as linear or ordinary memory, arises from the motion of unbound objects moving to infinity, such as during hyperbolic encounters~\cite{Zeldovich:1974gvh, Braginsky:1985vlg, Braginsky:1987}, or through asymmetric neutrino emission in supernovae~\cite{Vartanyan_2020}. The second type, non-linear memory, is driven by the energy flux of GWs themselves as they propagate to null infinity~\cite{Christodoulou:1991cr}. It is a prediction of the non-linear structure of GR and it is present in all GW sources. Because its source involves null GW radiation, it is also referred to as null memory, distinguishing it from the ordinary memory associated with the energy-momentum tensor of massive components.

The memory effect is intimately connected to BMS symmetries, which describe the symmetries of asymptotic space-times in General Relativity (GR). 
In particular, the conservation law associated with supertranslation symmetry provides an alternative formulation for computing the GW memory compared to the one presented here and in more detail in App.~\ref{app:memtheory}. Because this effect leaves a permanent displacement between two inertial observers, it is commonly referred to in recent literature as the displacement memory. However, this effect has been shown to represent only the leading contribution among several distinct types of memory, each linked to different asymptotic symmetries that preserve the structure of spacetime at null infinity~\cite{Flanagan:2018yzh,Grant:2021hga,Grant:2022bla,Siddhant:2024nft,Pasterski:2015tva,Nichols:2018qac,DeLuca:2024bpt}.
Furthermore, it is tied to the quantum structure of scattering amplitudes through soft theorems~\cite{Strominger:2014pwa,Strominger:2017zoo}. The detection of the GW memory effect is not only an additional confirmation of GR, but it is also relevant for its connection with the infrared structure of gravity and gauge theories. Indeed, asymptotic symmetries, soft theorems and memories represent the three different corners of the triangle diagram describing the infrared structure of gauge theories~\cite{Strominger:2014pwa,Strominger:2017zoo}.

When solving for the GW sourced by its energy-momentum tensor, the memory effect can be computed as an integral over the energy flux of the primary GWs \cite{Wiseman:1991ss,Thorne:1992sdb,Favata:2010zu,Heisenberg:2023prj}:
\begin{equation}\label{eq:memoryequation}
    \left[h^{\rm mem}_{ij}\right]^\text{TT}=\frac{4}{d}\int_{-\infty}^u \mathrm{d}u'\int \mathrm{d}\Omega' \frac{\mathrm{d}E}{\mathrm{d}u'\mathrm{d}\Omega'}\left[\frac{n'_i n'_j}{1 - n'_k N^k}\right]^\text{TT},
\end{equation}
where \( u \) is the retarded time, \( d \) the proper distance and \( n' \equiv n(\Omega') \) the unit radial vectors centered at the source. The TT superscript denotes the transverse-traceless projection along the line-of-sight direction \( N=n(\Omega)\), defined with the angular direction $\Omega$ in the source frame~\cite{Heisenberg:2023prj}. We summarize the derivation of Eq.~\eqref{eq:memoryequation} in App.~\ref{app:memtheory}.
The gauge-invariant energy flux of primary waves is defined as 
\begin{equation}\label{eq:energyflux}
    \frac{\mathrm{d}E}{\mathrm{d}u'\mathrm{d}\Omega'}=\frac{d^2}{16\pi}\langle \dot{h}_{+}^2+\dot{h}_{\times}^2\rangle,
\end{equation}
where $h_{+}$ and $h_{\times}$ are the amplitudes of the corresponding polarization states and the spacetime average $\langle...\rangle$ is to ensure a well-defined energy content of the GWs \cite{PhysRev.121.1556,PhysRev.166.1272,maggiore2008gravitational,Favata:2011qi,Heisenberg:2023prj}. 

Decomposing the GW strain, $h=h_+-ih_\times$ in spin-weighted spherical harmonics, as defined in Eq.~\eqref{eq:SW Memory quantity GR} in terms of $(\ell, m)$, 
the energy flux \( \propto |\dot{h}_{(2,2)}|^2 \) is largely dominated by the \((2,2)\) mode. Due to symmetry, this means the memory primarily appears in the \((2,0)\) mode, which in the reference frame of the source with the $z$-axis pointing in the direction of the observer, can be identified as the $+$ polarization,
as explained in App.~\ref{app:memtheory}.

\begin{figure}[t]
    \centering
    \includegraphics[width = \linewidth]{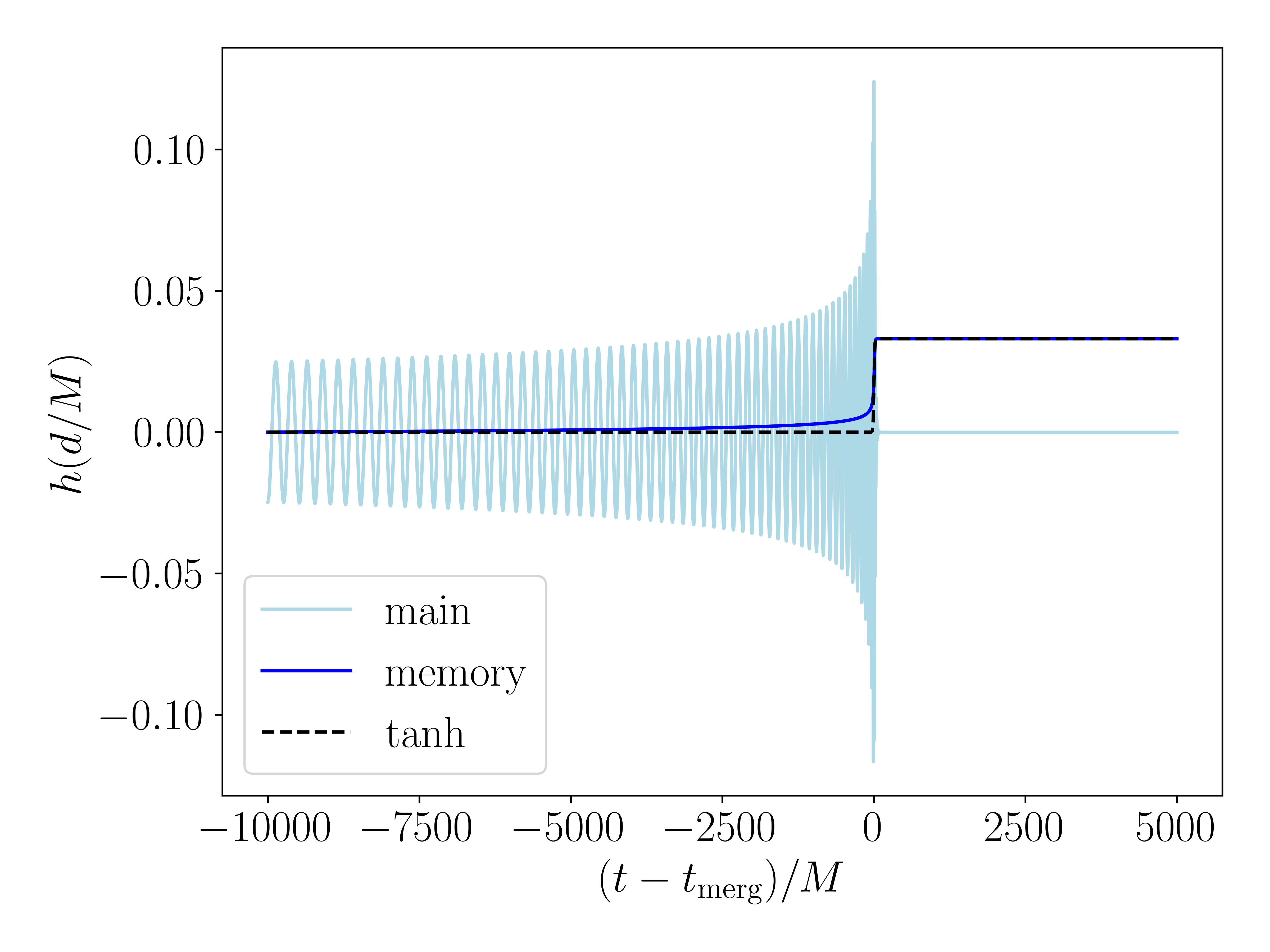}
    \caption{
    Comparison between the dominant oscillatory gravitational waveform \( ({\rm main} )\) associated with the (2,2) mode, and the memory waveform \( ({\rm mem}) \) for an edge-on system. For reference, we also include the analytical approximation of the memory from Eq.~\eqref{eq:memapproxTD}, using a characteristic rise time of \( \Delta\tau = 13M \). 
   } 
    \label{fig:IMMplot}
\end{figure}

A key feature of GW memory is that its value at any given time depends on the entire history of the binary—making it a hereditary effect. This results in a final amplitude comparable to the peak of the dominant oscillatory signal. From numerical waveforms,\footnote{We compute the memory from the public package \texttt{GWMemory}~\cite{PhysRevD.98.064031} which numerically evaluates Eq.~\eqref{eq:memoryequation} for given input waveforms. For the latter, we used the \texttt{NRHybSur3dq8} waveform model~\cite{PhysRevD.99.064045}.}
we find that the peak strain satisfies \( \max\{ h\} \simeq 3.7 \max\{ h_{\rm mem}\} \) for an edge-on system, where the memory is maximized. Fig.~\ref{fig:IMMplot} shows both the dominant oscillatory signal, \( h_{\rm main} \), which is the $(2,2)$ mode and the memory \( h_{\rm mem} \), which is the $(2,0)$ mode. The latter follows a characteristic step-like behavior, with the transition occurring around merger time \( t_{\rm m} \), when most of the energy is released.

\begin{figure}[t!]
    \centering
    \includegraphics[width = \linewidth]{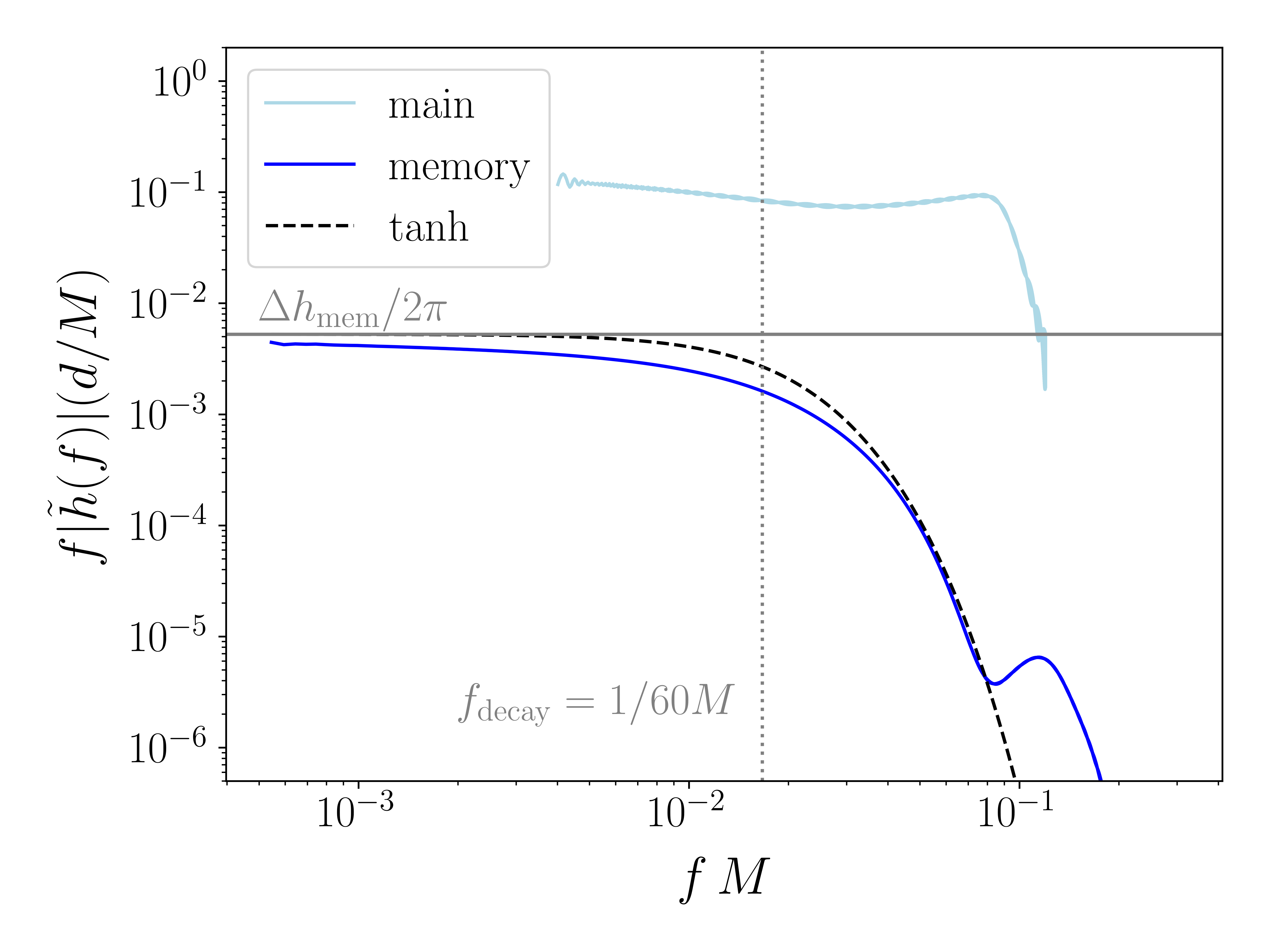}
    \caption{Characteristic strain in frequency domain for the system shown in Fig.~\ref{fig:IMMplot}, comparing the main oscillatory signal, the memory, and its analytical approximation. We find that Eq.~\eqref{eq:memapproxTD} with \(\Delta \tau \sim 13M \) provides a good fit to the memory signal, particularly in capturing the decay at frequencies above \( f_{\rm decay} = (60M)^{-1} \). At low frequencies, the characteristic strain approaches a constant value set by the final amplitude of the memory.
    }
   
    \label{fig:FTmem}
\end{figure}

A simple approximation for this behavior is given by  
\begin{equation}\label{eq:memapproxTD}
    h_{\rm tanh}(t) = \frac{\Delta h_{\rm mem}}{2} \left[\tanh{\left(\frac{- \tau}{\Delta\tau}\right)} +1\right],
\end{equation}
where $t_m-t \equiv \tau$ is the time to merger and
where \( \Delta \tau \) defines the rise time of the memory and $\Delta h_{\rm mem}$ its final amplitude.  We find that the difference between the final value and the initial value of the ${(2,0)}$ mode is $d/M \Delta h_{2,0}\simeq 0.08$ for a system of equal mass binary with total mass $M$ at distance $d$, which leads to\footnote{
This is somewhat lower than the value given in Ref.~\cite{Pollney_2011}, which obtained $d/M h_{2,0}=0.097\pm 0.002$. This can be traced back to the initial time inserted as a lower integration boundary in Eq.~\eqref{eq:memoryequation}, which in Ref.~\cite{Pollney_2011} was taken to be $-\infty$ whereas in this paper we focus on the memory produced around the merger time. Concretely, we consider an initial time $\tau_{\rm in}= 10^5 M$, capturing roughly the final 100 cycles, as shown in Fig.~\ref{fig:IMMplot}. Our results are robust to varying this choice. }
\begin{equation}\label{eq:Deltamem}
    \Delta h_{\rm mem}(\iota)\simeq 1.5\times 10^{-18} \sin^2{(\iota)}
    \left ( \frac{M}{M_\odot} \right )
    \left ( \frac{\rm kpc}{d} \right ),
\end{equation}
where the inclination angle $\iota$ is the angle between the observer and the normal vector to the binary plane.
The Fourier transform of Eq.~\eqref{eq:memapproxTD} then takes the form\footnote{We use the following convention for the FT and inverse FT \[ \Tilde{h}(\omega)= \int_{- \infty}^{\infty} {\rm d}t \, h(t) e^{-i \omega t} \quad \text{and}\quad h(t)=\frac{1}{2\pi}\int_{- \infty}^{\infty} {\rm d}\omega \, \Tilde{h}(\omega)e^{i\omega t} .\]}
\begin{align}\label{eq:FTmem}
    \Tilde{h}_{\rm tanh}(f) &= 
    \Delta h_{\rm mem} 
    \left[ 
    -\frac{ i\pi \Delta \tau }{2} 
    \sinh^{-1}{(\pi^2 \Delta \tau  f)} 
    +  i\delta(f) 
    \right ]
    \notag\\
    &\xrightarrow{f\ll \Delta \tau^{-1}} - i  \Delta h_{\rm mem} \left[ \frac{1}{2\pi f} - \delta(f) \right],
\end{align}
where the $\delta$-function accounts for a constant contribution in time ensuring $h_\text{mem}(t) = 0$ at $t \ll t_m$.
At frequencies lower than the inverse of the rise time, the signal follows a \( f^{-1} \) scaling, characteristic of a step function. Frequencies above \( \Delta \tau^{-1} \) are suppressed, and we find that  \( \Delta \tau \sim 13 M \) provides a good fit to the full expression, which features a decay setting in at  \( f_{\rm decay} \sim (60 M)^{-1} \)~\cite{Gasparotto:2023fcg} (see Fig.~\ref{fig:FTmem}).
Notably, while the amplitude of the memory signal is comparable to the primary signal in the time domain, its contribution to the characteristic strain in frequency space remains subdominant. This can be understood from the fact that the primary signal is quasi-monochromatic at low frequencies, thus spending a lot of time and accumulating many cycles per frequency bin at low frequencies. In contrast, memory shows the characteristic burst-like behavior of being concentrated in time but distributed in frequency. Consequently, the total energy released must be integrated over a considerably broader range of frequencies in comparison to the dominant signal.

\subsection{Memory signal template}
As discussed in the previous section, due to the characteristic low-frequency behavior of the memory, it is, in principle, possible to detect a merger occurring at frequencies above the detector's sensitivity range. To our knowledge, the only searches for out-of-band mergers using real data have been conducted in Ref.~\cite{Agazie:2025oug} (see also \cite{NANOGrav:2019vto,NANOGrav:2023vfo}) in the context of pulsar timing array GW searches, and the authors of~\cite{Ebersold:2020zah}, who searched for subsolar mergers in the second observing run (O2) of the LIGO and Virgo detectors. Using LIGO/Virgo as triggers to search for out-of-band memory signals in LISA was proposed in~\cite{Ghosh:2023rbe}.
In this regime, the rise time of the memory corresponds to frequencies higher than the sensitivity range of the instrument, which is primarily sensitive to the $f^{-1}$ component of the signal.
Modeling the detector as a high-pass filter motivated by the rapid drop-off in sensitivity below a frequency $f_{\rm min}^{\rm det}$,
Ref.~\cite{Ebersold:2020zah} employed the unmodeled search method \textit{coherent WaveBursts} (cWB)~\cite{Klimenko:2008fu,Klimenko:2015ypf}, which, while efficient and model-independent, is in general not as sensitive as an optimal matched filtering search.\footnote{By comparing the sensitivity of the cWB search in Ref.~\cite{Ebersold:2020zah} with our estimates, we find a potential sensitivity gain of two orders of magnitude. There are however caveats to this comparison, most notably that an actual search on real data faces challenges which are not accounted for in theoretical sensitivity estimates. Moreover, Ref.~\cite{Ebersold:2020zah} was performed on O2 data whereas our estimate is based on design sensitivity, see App.~\ref{app:sens_curves}.
}
Given the simplicity and universality of the memory waveform in this regime, 
we argue that incorporating a memory-based template search into current and future analyses is not only feasible but essential because it allows probing weak events which otherwise would not pass the detection threshold of an unmodeled search.

 To this end, let us consider the memory signal $s(t)$ as seen by the detector, which is described by the convolution of the detector response function $R({\bm k})$ capturing the detector geometry with the gravitational wave,
\begin{align} \label{eq:s}
 s(t) = \int_{- \infty}^\infty df R(\bm k) \tilde h(f) e^{2 \pi i f t } \,.
\end{align}
Here ${\bm k} =  2 \pi f  \hat{\bm k}$ denotes the wave vector of the GW in the detector frame for a GW originating from a direction $- {\hat{\bm k}}$ in the sky. Assuming that the primary signal peaks significantly out-of band, $f \ll \Delta \tau^{-1}$, and that we can model the detector response as a flat response in frequency limited to a frequency window $[f_{\rm min}^{\rm det}, f_{\rm max}^{\rm det}]$, inserting \eqref{eq:FTmem} yields,
\begin{align}\label{eq:HPmem}
 s(t) & \simeq R(\hat{\bm k}) \Delta h_\text{mem} \left(\int_{f_{\rm min}^{\rm det}}^{f_{\rm max}^{\rm det}} \frac{(e^{2 \pi i f t}- e^{- 2 \pi i f t})}{2 \pi i f}df + \frac{1}{2}\right) \nonumber \\
 & = R(\hat{\bm k}) \frac{\Delta h_\text{mem}}{\pi} \left(\left [\text{Si}(2 \pi  f_{\rm max}^{\rm det} t)
-\text{Si}(2 \pi  f_{\rm min}^{\rm det} t) \right ] + \frac{\pi}{2} \right)
\end{align}
where ${\rm Si}$ is the sine integral function.
If we moreover assume a large hierarchy between the observable frequencies, $f_{\rm max}^{\rm det} \gg f_{\rm min}^{\rm det}$,
Eq.~\eqref{eq:HPmem} becomes
\begin{align}
 s(t)&=
   R(\hat{\bm k}) \Delta h_{\rm mem}\left [\frac{1}{2} +
  \text{sign}(t) - \frac{2}{\pi } \text{Si}(2  \pi f_{\rm min}^{\rm det}  t)
  \right .
  \nonumber \\
  & \left . 
  - 
  \left (\frac{f_{\rm min}^{\rm det}}{f_{\rm max}^{\rm det}} \right )
  \frac{\cos (2 \pi f_{\rm max}^{\rm det} t  )}{\pi ^2 f_{\rm min}^{\rm det} t
  }
  +{\cal O}
  \left ( \frac{{f_{\rm min}^{\rm det}}^2}{{f_{\rm max}^{\rm det}}^2} \right)
  \right].
  \label{eq:s_approx}
\end{align}
In this limit, the memory signal takes a universal form, depending only on the detector properties, the merger time, and one free parameter describing the effective amplitude $R(\hat{\bm k}) \Delta h_\text{mem}$. The latter captures all parameters of the PBH binary, including the sky position. More precise signal templates can be obtained by maintaining the full frequency and angular dependence of the response function, and we provide these explicitly in the App.~\ref{app:response} for the case of LIGO and LISA. In these cases, we find a mild dependence of the signal template on the sky position, arising from the frequency dependence of the response function.

\begin{figure}[t]
    \centering
    \includegraphics[width=\linewidth]{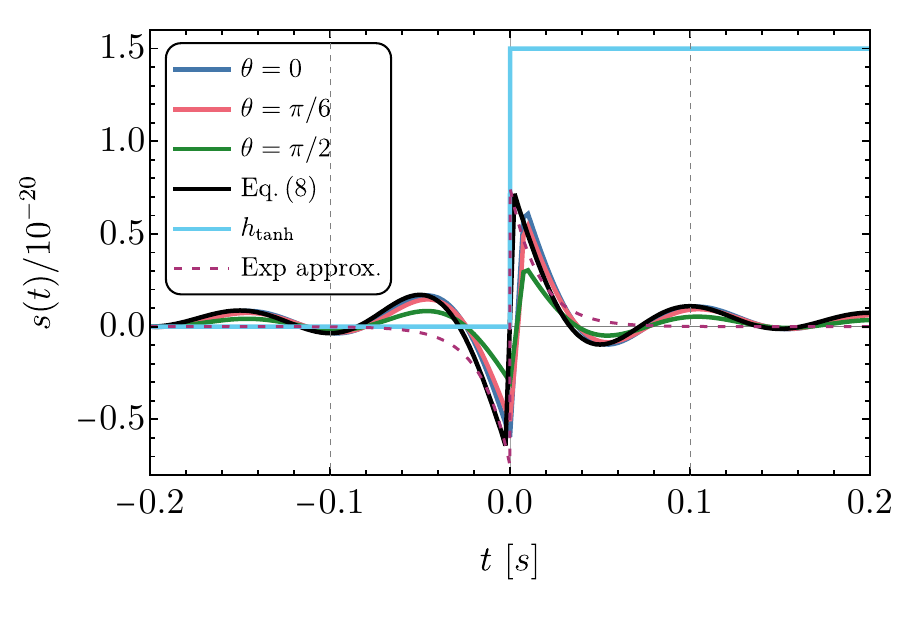}
    \caption{
    Time-domain memory event from a merger of an equal-mass binary with total mass $M=10^{-2} M_\odot$ at distance $d=1$ kpc in LIGO. The cyan curve shows the memory signal as given by Eq.~\eqref{eq:memapproxTD} while the black curve shows leading order result for the signal template compared to the full result for three different sky positions (in blue, pink and green).
    The dashed purple line indicates the exponential suppression $e^{-2 \pi f_{\rm min}^{\rm det} t}$.
    }
    \label{fig:bandpassed memory}
\end{figure}
To illustrate this, Fig.~\ref{fig:bandpassed memory} compares the leading order terms of Eq.~\eqref{eq:s_approx} [black] with the full result in Eq.~\eqref{eq:s} [colored] for the case of an equal-mass binary with total mass $M=10^{-2} M_\odot$ at a distance of $d=1$ kpc in LIGO, assuming $f_{\rm min}^{\rm det} = 10$ Hz, as in Ref.~\cite{Ebersold:2020zah}. Accounting for the full response function reveals that the time scale of the characteristic rapid change in the signal at $t = t_m$ is set by the effective interferometer arm length,  which is about 4~ms for LIGO. (For the case of $f \ll \Delta \tau^{-1}$, otherwise the latter time scale would also enter.) This can be understood intuitively as the memory signal Eq.~\eqref{eq:memapproxTD} changing the effective interferometer arm length.
The slower oscillations are induced by the high-pass filter in Eq.~\eqref{eq:HPmem}, and consequently, their frequency is set by $f_{\rm min}^{\rm det}$. The latter is given by either the (uninterrupted) observation time of the detector or the high-pass filters applied in the analysis (as done in particular in most LIGO analyses to remove the frequency range dominated by seismic noise).\footnote{
We caution that the oscillatory features associated with $f_{\rm min}^{\rm det}$ are induced by the data processing and hence by construction are present in the data even in the absence of a memory signal. They should hence not be used for signal search.
}
One such period, corresponding to $1/f_{\rm min}^{\rm det} = 0.1$ s, is highlighted by the dashed gray vertical lines in Fig.~\ref{fig:bandpassed memory}. The amplitude of the band passed memory is exponentially suppressed as $e^{-2\pi f_{\rm min}^{\rm det} t}$ as shown with the purple dotted lines.

In the limit of out-of-band sources, the characteristic memory signal is primarily determined by detector properties. For instance,  for LISA, the approximation of a frequency independent response is no longer appropriate, and the use of time-delayed interferometry, required to mitigate the laser noise, eliminates the slow oscillations, leaving only the sharp feature at $t = t_m$, see Ref.~\cite{Ghosh:2023rbe} and App.~\ref{app:response}.\footnote{The case of the LISA response to the memory signal from in-band binary mergers is discussed in Ref~\cite{Inchauspe:2024ibs}, together with the detection prospects for the massive black hole population.
}
As the results above demonstrate, the signal template can in general only depend on four parameters: the merger time, an effective amplitude and two angles for the sky position. While this universality makes a matched filtering search very efficient, the reverse side of the coin is an increased risk of noise at these frequencies since they are set by detector properties and not the GW signal.  A coincident detection, either in the same frequency band or in different bands, thus seems crucial to claim a detection.

\subsection{GWs signals from inspiraling binaries}\label{subsec:GWinspiral}

To describe the inspiral signal, we adopt the frequency domain GW strain produced by a circular BH binary at a given frequency $f$ at zeroth post-Newtonian (0-PN) order \cite{Maggiore:1900zz,Antelis:2018sfj}
\begin{align}
   \tilde  h_\lambda(f) &= \bigg(\frac{5}{24}\bigg)^{1/2} \frac{1}{\pi^{2/3}} \frac{1}{d}
            {\cal M}^{5/6} f^{-7/6} e^{i \psi} Q_\lambda(\iota)
    \label{eq:strainPBHs}
\end{align}
with $Q_+(\iota) = (1 + \cos^2 \iota)/2$ and $Q_\times = \cos \iota$; 
$G$ is Newton's constant; ${\cal M} \equiv (m_1 m_2)^{3/5}/(m_1+m_2)^{1/5}$ is the chirp mass of a binary with constituent masses $m_1$, $m_2$;  $d$ is the proper distance from the binary to the observer and $\psi$ denotes the phase of the binary.
We assume the GW emitted by the inspiral (being twice the orbital frequency) is described by Eq.~\eqref{eq:strainPBHs} up to the Innermost Stable Circular Orbit (ISCO) \cite{Maggiore:2007ulw}
\begin{equation}
    f_{\rm ISCO}
    =  4.4 {\rm kHz} \left ( \frac{M_\odot}{m_1+m_2} \right ).
\end{equation}

We can compute the SNR by using Eq.~\eqref{eq:SNRdef},
where the limits are selected in such a way that the signal is integrated over the detector sensitivity band, and accounting for the inspiral intrinsic evolution during the observation time $T_{\rm obs}$.
Therefore, we set the detector frame integration limits to
\begin{align}
    f_1 &= {\rm max} [f_{\rm in}, f_{\rm min}^{\rm det}]
    \\
    f_2 &=  {\rm min} [f_{\rm ISCO},f_{\rm max}^{\rm det}, f_{\rm max}^{\rm T_{\rm obs}}(f_{\rm in})]. 
\end{align}
The meaning of each frequency is described in Tab.~\ref{tab:relfreq}.

We define $f_{\rm in}$ as the GW frequency recorded by the observer in the detector frame at the start of the observation.
We should only consider $f_{\rm in}$ in the range $f_{\rm in} \in [f_{\rm min}^{\rm T_{\rm obs}}, f_{\rm ISCO} ]$ and masses for which $f_{\rm ISCO} > f_{\rm min}^{\rm det}$.
The former selects signals that enter the detectable frequency range, while staying below the maximal frequency excited by the system. The latter condition limits the systems to the ones that are light enough to excite observable frequencies. 
Notice that all the above is defined in the detector frame.
Then, frequencies $f_{\rm min}^{\rm T_{\rm obs}}$ and $f_{\rm max}^{\rm T_{\rm obs}}(f_{\rm in})$ are computed as
\begin{align}
f_{\rm min}^{T_{\rm obs}}&=f_{\rm coal}(\tau(f^{\rm det}_{\rm min})+ T_{\rm obs}),
\\
f_{\rm max}^{T_{\rm obs}}&=f_{\rm coal}(\tau(f_{\rm in})- T_{\rm obs}),
\end{align}
where the GW frequency at a given time before coalescence is given by (neglecting corrections induced by eccentricity) 
\begin{align}\label{eq:f(tau)}
    f_{\rm coal}(\tau) &= 
    \frac{5^{3/8}}{8 \pi}
    \tau^{-3/8}{\cal M}^{-5/8}
    \nonumber \\
    &
    = 0.23\, {\rm Hz}
    \left ( \frac{{\rm yr}}{\tau}\right )^{3/8}
    \left ( \frac{M_\odot}{m_1}\right ) ^{5/8}
    \left ( \frac{1+q}{q^3} \right )^{1/8}.
\end{align}
Conversely, the time of coalescence in terms of the GW frequency $\tau(f)$ is simply the inverse of Eq.~\eqref{eq:f(tau)}.

{
\renewcommand{\arraystretch}{1.4}
\setlength{\tabcolsep}{4pt}
\begin{table}[t]
\begin{tabularx}{\columnwidth}{|c|l|}
\hline
\hline
\multicolumn{2}{|c|}{Detector} 
\\
\hline
$f_{\rm min}^{\rm det}$ &
Min detector frequency
\\
\hline
$f_{\rm max}^{\rm det}$ &
Max detector frequency
\\
\hline
\hline
\multicolumn{2}{|c|}{Inspirals} 
\\
\hline
$f_{\rm in}$ &
Frequency at the start of the observation
\\
\hline
$f_{\rm ISCO}$ &
Max frequency of
the binary's inspiral
\\
\hline
$f_{\rm min}^{\rm T_{\rm obs}}$ &
Min frequency leading to $f_{\rm min}^{\rm det}$ in a time $T_{\rm obs}$
\\
\hline
$f_{\rm max}^{\rm T_{\rm obs}}(f_{\rm in})$ &
Max frequency reached from $f_{\rm in}$ in a time $T_{\rm obs}$
\\
\hline
\hline
\multicolumn{2}{|c|}{Memory} 
\\
\hline
$f_{\rm decay}$ & Max frequency before memory signal decay
\\
\hline
\hline
\end{tabularx}
\caption{ 
Relevant frequencies for computing the SNR.
}
\label{tab:relfreq}
\end{table}
}

\subsection{Signal-to-noise ratio}

Given the existence of a simple signal template for the memory,
we compute the detector’s sensitivity to the memory and the dominant signals by evaluating the signal-to-noise ratio (SNR) in the context of a matched filtering search. The SNR is given by 
\begin{align}
\hat{\rho}^2 =
4 \int_{f_1}^{f_2} {\rm d} f  \frac{|\tilde{s} (f)|^2}{P_n(f)}=4 \int_{f_1}^{f_2} {\rm d} f  \frac{|\tilde{h} (f)|^2}{S_n(f)} , \label{eq:SNRdef}
\end{align}
where $|\tilde{h}|^2=\tilde{h}_\times^2+ \tilde{h}_+^2$ and $S_n=P_n/R^2$ is the noise-equivalent strain power spectral density of the detector~\cite{Robson:2018ifk}. The response functions depend on the direction of the source, we report the averaged value over all sky directions and the sensitivity curves used in App.~\ref{app:sens_curves}. We also average over the inclination angle for both the primary and the memory signal.
The integration limits $f_1$ and $f_2$ are set by the minimum and maximum frequencies of the detector or by the frequency range of the signal over the observation period $T_{\rm obs}$
as discussed in detail for the dominant GW signal in the previous section.
We stress that, for distant sources merging at a redshift $z\gtrsim {\cal O}(1)$, cosmological redshift becomes important, and the integration in Eq.~\eqref{eq:SNRdef} is defined in the detector frame.

\section{Event rates}\label{sec:PBHmergerrate}

While all the considerations above describe the properties of the GW signatures of a generic binary BH system, in this section we focus on the specific case of PBH binaries. This is because only the primordial scenario, at odds with standard astrophysical formation mechanism, is able to produce subsolar mass BHs \cite{1931ApJ....74...81C}, of particular interest to our discussion. 
Needless to say, the detection of a subsolar mass compact object could suggest either a previously unknown formation mechanism that extends beyond standard-model stellar core-collapse scenarios~\cite{Metzger:2024ujc,Muller:2024aod} or provides evidence for new physics. Beyond PBHs, one could also consider other exotic objects~\cite{Cardoso:2019rvt}, such as Q-balls and boson stars~\cite{Coleman:1985ki,Colpi:1986ye, Liebling:2012fv}, as well as fermion-soliton stars~\cite{Lee:1986tr,DelGrosso:2023trq,DelGrosso:2023dmv}. In either case, such a discovery would have profound implications. We will focus on the PBH scenario, leaving a thorough assessment of the merger rate of exotic alternatives for future work.  

\subsection{PBH merger rates}\label{sec:PBH merger rate}
PBHs form at very high redshift and can pair in binaries already before matter-radiation equality.
As shown in the literature, see Ref.~\cite{Raidal:2024bmm} for a recent review,
the dominant channel contributing to the present-day merger rate is due to binary formation after decoupling from the Hubble flow in the early universe. We consider this case and comment about alternative binary formation channels below.

The merger rate density is given by \cite{Raidal:2018bbj}
\begin{align}
 \frac{{\rm d}^2 {\cal R} (t,d) }{{\rm d} \ln m_1{\rm d} \ln m_2} 
 &= 
\left(\frac{0.038}{\rm kpc^3 \, yr} \right)
[1 + \delta (d)]
f_{\rm PBH}^{\frac{53}{37}} 
\left ( \frac{t}{t_0} \right )^{-\frac{34}{37}}   
 \nonumber \\
 & \times 
 \left[ \frac{m_1 m_2}{(m_1+m_2)^2} \right] ^{-\frac{34}{37}} 
 \left ( \frac{m_1+m_2}{10^{-12}M_\odot} \right )^{-\frac{32}{37}}  
\nonumber \\
 & \times 
 S\left (t, M, f_{\rm PBH}, \psi  \right )
\psi(m_1) \psi(m_2)
 ,
\label{eq:PBHrate}
\end{align}
where we introduced the current age of the Universe $t_0 = 13.8\,  {\rm Gyr}$
and the PBH mass distribution $\psi (m)\equiv (m/\rho_{\rm PBH}) {\rm d} n_{\rm PBH} /{\rm d} \ln m$,
normalized such that 
$\int {\rm d} \ln m \psi (m) = 1$.
As standard in the PBH literature, we introduced the PBH abundance $f_{\rm PBH} \equiv \Omega_{\rm PBH}/\Omega_{\rm DM}$. It should be considered a proxy for the PBH number density $n_{\rm PBH} \sim f_{\rm PBH} \rho_{\rm DM} / \langle m \rangle $.

The suppression factor $S\left ( M, f_{\rm PBH},\psi \right )$
includes the effect of binary interactions with the surrounding environment in both the early- and late-time Universe.
This can be written as the product of two separate contributions 
$S\equiv S_{\rm E} \times S_{\rm L} $~\cite{Hutsi:2020sol}, where
\begin{align}
S_{\rm E} 
 & \approx 1.42 
 \, e^{-  \bar N }
 \left ( \frac{\langle m^2 \rangle/\langle m\rangle^2}{\bar N +C} + \frac{\sigma ^2_\text{\tiny M}}{f^2_{\rm PBH}}\right ) ^{-21/74}  ,
\label{S1}
\end{align}
reduces the PBH merger rate due to interactions close to the formation epoch between the forming binary and both the surrounding DM inhomogeneities with characteristic variance $\sigma_\text{\tiny M}^2 \simeq 3.6 \times 10^{-5}$ as well as neighboring PBHs~\cite{Ali-Haimoud:2017rtz,Raidal:2018bbj,Liu:2018ess}.
The characteristic number of PBHs around the binary is defined as\footnote{The explicit expression for the constant $C$ entering in Eq.~\eqref{S1} can be found in Ref.~\cite{Hutsi:2020sol}. We omit it here for brevity.}
\begin{equation}
\bar N 
\equiv 
\frac{(m_1+m_2)}{\langle m \rangle } 
\frac{f_{\rm PBH}}{f_{\rm PBH}+ \sigma_\text{\tiny M}}.
\end{equation}
The second contribution
\begin{align}\label{eq:supS2}
S_{\rm L} & \approx 
	\text{min} 
	\left [ 1\, ,\, 9.6 \times 10^{-3} x ^{-0.65} \exp \left ( 0.03 \ln^2 x \right )  \right ] ,
\end{align}
with $x \equiv (t(z)/t_0)^{0.44} f_{\rm PBH}$, accounts for the effect of successive disruption of binaries that populate PBH clusters formed from the initial Poisson inhomogeneities,
conservatively estimated assuming that the entire fraction of binaries included in dense environments is disrupted.\footnote{We will only consider standard formation scenarios for which initial clustering is negligible \cite{Crescimbeni:2025ywm}, and PBH structure formation proceeds through Poisson induced inhomogeneities \cite{Inman:2019wvr,DeLuca:2020jug}.}

The rate shown in Eq.~\eqref{eq:PBHrate} is boosted when considering small distances from the Earth due to the galactic overdensity. Assuming the distribution of PBH binaries to follow the one of the dark matter, the local merger rate density is enhanced by the overall factor $[1 + \delta (d)] $ included in Eq.~\eqref{eq:PBHrate} \cite{Pujolas:2021yaw}.
This gives the differential merger rate density dependence on the distance from the earth $d$.
We defined the overdensity factor $\delta(r) \equiv \rho_\text{\tiny DM}(r)/\bar \rho_\text{\tiny DM}$. 
The Milky Way dark matter halo is modeled as a Navarro-Frenk-White density profile,
$\rho_\text{\tiny{DM}}(r) = {\rho_0} r_0/ [r \left(1+{r}/{r_0}\right)^{\!2} ] \,$, with the normalization $\rho_0$ fixed such that 
$\rho_\text{\tiny DM}(r=r_\odot) = 7.9\times 10^{-3}M_\odot/{\rm pc}^3$ at the 
the solar system location is $r_\odot \simeq  8.0\,{\rm kpc}$, and  $r_0 = 15.6\, {\rm kpc}$.
We can simplify the overdensity within a volume of radius $d$ around the Earth location as
\begin{align}\label{eq:delta(r)over}
\delta(d)\equiv
\rho_\text{\tiny DM}(d)/{\bar \rho_\text{\tiny DM}}=
\begin{cases}
\rho_\text{\tiny DM}(r_\odot)/{\bar \rho_\text{\tiny DM}}, &
\quad  d<r_\odot,	
\\
\rho_\text{\tiny DM}(d) /{\bar \rho_\text{\tiny DM}}, &
\quad   d \gtrsim  r_\odot.
\end{cases}
\end{align}

As we will see, a crucial property of early universe (EU) formation scenarios is that they produce binaries sufficiently wide to inspiral over a timescale comparable to or greater than the current age of the universe. Consequently, there exists a large population of inspiraling sources emitting GWs at frequencies $f \ll f_{\rm ISCO}$. This population plays a key role in constraining the abundance of light PBHs.
However, it is important to emphasize that alternative formation scenarios may lead to different conclusions. For instance, if binary formation proceeds through gravitational capture (cap) \cite{1989ApJ...343..725Q,Mouri:2002mc} in the late-time universe, the merger timescale is significantly shorter. In this scenario, the two objects form a hard and highly eccentric binary that merges promptly, typically within only a few orbital cycles. The maximum coalescence time is approximately
$
    \tau \approx 0.1~{\rm Myr} \left( {m}/{M_\odot} \right)
$
(assuming typical velocity dispersion in PBH clusters) \cite{OLeary:2008myb}, implying that sufficiently low GW frequencies are not emitted.

Additionally, three-body interactions (3b) can also lead to PBH binary formation \cite{Korol:2019jud,Kritos:2020wcl,Franciolini:2022ewd}. This process occurs at sufficiently high redshifts in PBH clusters induced by Poissonian initial noise. This channel tends to produce relatively wider binaries compared to capture, as sufficient energy to bind PBH pairs is extracted by the escaped third object. Therefore, binaries formed through this channel are expected to inspiral from very low frequencies. 

In conclusion, while these alternative binary formation channels are \textit{possible}, they remain subdominant within the standard scenario \cite{Ali-Haimoud:2017rtz,Franciolini:2022ewd,Raidal:2024bmm}. This is particularly true for light PBHs with masses below the stellar range, as the respective event rates scale as follows:
$
    {\cal R}^{\text{EU}} \propto m^{-32/37}, 
    {\cal R}^{\text{cap}} \propto m^{-11/21}, 
    {\cal R}^{\text{3b}} \propto m^{-(11 -- 16)/21}.
$

\subsection{Number of detectable events}
\label{subsec:Ndet}

In this section, we compute the number of detectable events at GW experiments as a function of the PBH population parameters $f_{\rm PBH}$ and $m_{\rm PBH}$.
For presentation purposes, we assume a narrow mass distribution and thus $\langle m ^n\rangle = m^n$.
To include both mergers as well as binaries in the early stages of their inspiral, which appear as nearly monochromatic signals in the GW detectors, we start by defining the number density of binaries per frequency bin $\d n_{b}/\d f$.

\begin{figure}[t!]
    \centering
    \includegraphics[width = \linewidth]{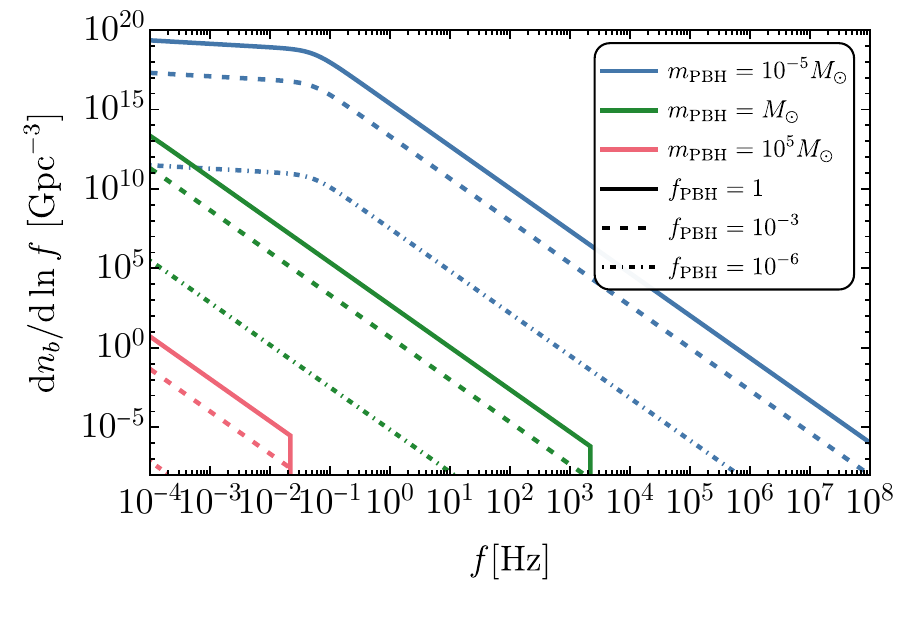}
    \caption{
    The differential comoving number density of binaries today as a function of frequency for representative values of $f_{\rm PBH}$ and $m_{\rm PBH}$. The sharp cut corresponds to the ISCO frequency. 
    The large frequency power-law $\propto f^{-8/3}$ is induced by $f_{\rm coal}^\prime$, while the break at lower frequencies indicates where $\tau(f)$ becomes comparable to $t_0$ and the additional scaling in the merger rate density leads to the number density scales as $\propto f^{-8/37}$.
   } 
    \label{fig:dn/df}
\end{figure}

We consider binaries in any phase of their inspiral. 
At the time $t_0$, the comoving number density of binaries that would merge in the interval  $[t_0+\tau,t_0+ \tau + d\tau]$ is given by ${\cal R}(t_0 + \tau,d) {\rm d} \tau$.
Assuming GW emission has sufficiently circularized their orbits, i.e. eccentricity is negligible,\footnote{This is a good approximation for binaries formed in the early universe, as GW driven binary evolution is expected to have circularised the orbit well before $t_0$ \cite{Franciolini:2021xbq}.
}
the frequency of GWs emitted by a binary with a  time to coalescence $\tau$
is given by Eq. \eqref{eq:f(tau)}.
Then, the number density of sources at a time $t$ emitting at a frequency $f$ can be calculated as 
\begin{equation}\label{eq:dndf}  
\frac{\d n_b(t, d,f )}{\d f}  
= 
\left |\frac{\d f_{\rm coal}(\tau(f))}{{\rm d} \tau}\right|^{-1} 
{\cal R}(t + \tau(f),d)\,.
\end{equation}  
In \Cref{fig:dn/df}, we show the number density of PBH binaries as defined in Eq.~\eqref{eq:dndf} for representative values of both $f_{\rm PBH}$ and $m_{\rm PBH}$. The decrease as a function of frequency reflects the frequency evolution for fixed PBH masses, which speeds up at higher frequencies following the characteristic `chirp' signature. Consequently, at higher frequencies, a given binary spends less time per frequency bin, and thus the number density at a given frequency decreases. For nearly monochromatic sources, such a steep decline is mitigated by the fact that the PBH population of binaries peaks at merger timescales much larger than the age of the universe.

The number of detectable merger events is computed by integrating the differential number density of objects as a function of the GW frequency at the start of the observations over the detectable volume. This means (e.g. \cite{Pujolas:2021yaw})
\begin{align}\label{eq:ndetdef}
    N_{\rm det} =
   \int \d f_{\rm in}
   \int \d z  
   \frac{1}{(1+z)}
   \frac{\d V_c}{\d z}
   \frac{\d n_b(t(z),d(z),f_{\rm in})}{\d f_{\rm in}}
   \nonumber 
   \\
\times \Theta( \hat \rho(f_{\rm in}, z)- \hat\rho_{\rm th}),
\end{align}
where $V_c$ is the comoving volume and $\hat \rho$ identifies the SNR as defined in Eq.~\eqref{eq:SNRdef}, applied to memory or the inspiral signal, respectively. The Heaviside step function in Eq.~\eqref{eq:ndetdef} introduces the selection bias based on which binary system would have SNR bigger than the threshold, customary set to $\hat{\rho}_{\rm th}=8$.
This, in practice, corresponds to prescribing the observable volume, or equivalently the maximum redshift, up to which binaries with mass $m$ inspiraling at $f_{\rm in}$ would be detectable.

Notice that, when considering sources at cosmological distances, the SNR is computed by integrating the GW signal in frequency in the detector frame, while 
we perform the integration on 
$f_{\rm in}$ in Eq.~\eqref{eq:ndetdef} 
in the source frame. The two are simply related by $f_d = f_s /(1+z)$.
When computing the SNR, this can be equivalently implemented for both the inspiral and memory signal by following the prescription of replacing $f_s \mapsto f_d$, $d \mapsto d_L = (1 + z) d$ and ${\cal M} \mapsto {\cal M}_z \equiv (1 + z) {\cal M}$.
At cosmological scales, the luminosity distance $d_L$ is redshift-dependent, and defined as
\be
\label{eq:dLz}
d_L(z) = \frac{1}{H_0} (1+z) \int_0^z \frac{dz'}{\sqrt{\Omega_M (1+z')^3 + \Omega_{\Lambda}}},
\ee
where
$H_0=100 h \, {\rm (km/s)/Mpc}$. We use the cosmological parameters given in Table 4, column 3 of \cite{Planck:2015fie}: $h=0.6790$, $\Omega_M=0.3065$, and $\Omega_{\Lambda}=0.6935$.  

When considering the memory signal, we focus directly on binaries close to the merger, where most of the memory signal is sourced. This means, we only consider a binary system for which $f_{\rm in}$ is such that the merger timescale is much smaller than the age of the universe. Combining Eqs.~\eqref{eq:dndf} and \eqref{eq:ndetdef} in the limit $\tau (f_{\rm in}) \ll t_0$, one can show that the integration over $f_{\rm in}$ turns into an integration over the observation time, leading to the familiar expression
    \begin{align}\label{eq:ndetdef_rate}
    N_{\rm det} =
   T_{\rm obs}
   \int \d z  
   \frac{\d V_c}{\d z}
   \frac{{\cal R} (t(z),d_L(z))}{(1+z)} \Theta( \hat \rho(f_{\rm in}, z)- \hat\rho_{\rm th}).
\end{align}

\begin{figure*}
    \centering
    \includegraphics[width = \linewidth]{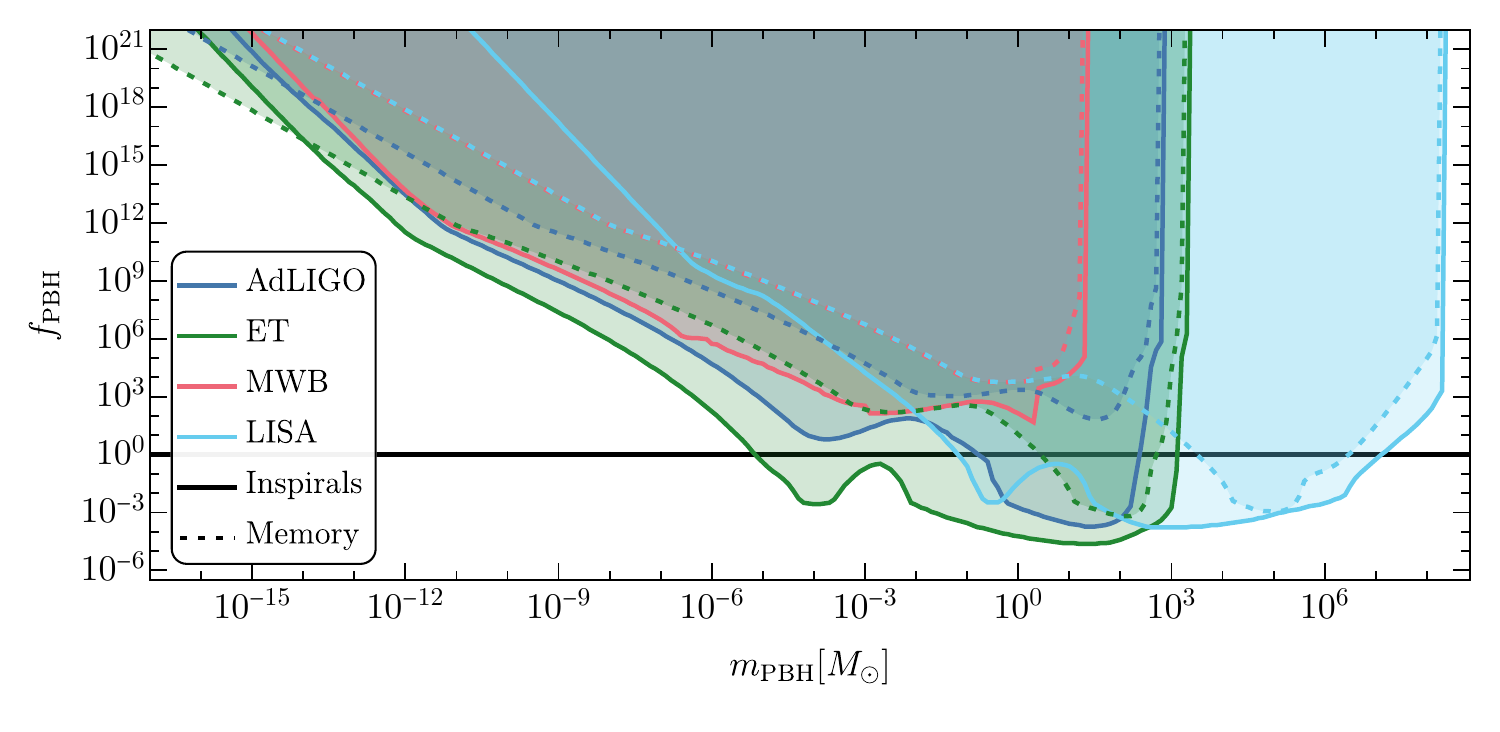}
    \caption{
    Sensitivity to PBH abundance corresponding to $N_{\rm det} > 1$ in 1 yr observation at the GW experiments AdLIGO (blue), ET (green), Magnetic Weber Bars (red), and LISA (cyan). The solid curves refer to the primary signal (inspirals and mergers), while the dashed curves are based on the memory signal. In the physical parameter space of $f_\text{PBH} < 1$, the strongest constraints are derived from the primary signal for all detectors considered. For out-of-band mergers, this implies that the GW signal generated in the early inspiral stage leads to a stronger signal than the memory effect.
   } 
    \label{fig:plots}
\end{figure*}

\section{Results}\label{sec:res}

We are now in the position to compare the reach of future detectors when searching for PBH binaries through inspiral or memory GW signatures. 
We consider different future GW experiments: 
Advanced LIGO (AdLIGO) at design sensitivity \cite{KAGRA:2013rdx},
Einstein Telescope (ET) \cite{Hild:2008ng,Punturo:2010zz,Hild:2010id}, 
Laser interferometer space antenna (LISA) \cite{LISA:2017pwj,LISACosmologyWorkingGroup:2022jok}, 
and the proposal of HFGW Magnetic Weber Bars (MWB)~\cite{Domcke:2024mfu}.
We give more details on the sensitivity of these detectors in App.~\ref{app:sens_curves}.

Fig.~\ref{fig:plots} shows the minimum abundance required to predict more than one detection ($N_{\rm det} > 1$) within a one-year observation period $T_{\rm obs} = 1 {\rm yr}$ across various GW experiments. Solid lines represent inspiral signals, while dashed lines correspond to memory signals. 
In the absence of detections, this line turns into a future bound on $f_{\rm PBH}$. 

We do not show other current and future constraints on this parameter space, e.g. derived from direct PBH searches through microlensing or indirect through accretion bounds, while we refer to \cite{Carr:2020gox} for a comprehensive list.
In Fig.~\ref{fig:plots} we extend the plot to values of the abundance $f_{\rm PBH} >1$. This unphysical parameter space is not meant to indicate realistic detection prospects, but just aims to compare the reach of different detection strategies, i.e. memory vs primary signals,
and to illustrate the relevant relative scalings of the different curves, even if irrelevant for current and near-future detectors.
Values $f_{\rm PBH} >1$ indicate that at this mass range, signals are only visible up to distances much below the ones required for discovery for a realistic PBH merger rate.

Let us focus first on the primary signal (inspiral and/or merger), as depicted by the solid curves.
All detectors except MWB feature a double peak structure. 
The strongest constraint, peaking at $m_{\rm PBH} \sim 10^2 M_\odot$ for both AdLIGO and ET corresponds to when the corresponding ISCO frequency sits at the minimum of the sensitivity curve.
For LISA instead, the minimum appears at $m_{\rm PBH} \sim 10^3 M_\odot$ (see below for a more detailed discussion).
At smaller masses, we see the appearance of a second feature, where the bound strengthens again. This is due to the local DM overdensity boosting the rate of events of systems we can observe up to a few tens of kpc. For larger observable volumes, this local rate enhancement is washed out. 
For MWB this effect is not present, as the sensitivity is always limited to events that are within our galaxy. The spiky feature in the MWB lines is due to the resonant dip in the sensitivity curve shown in Fig.~\ref{fig:psd}.

To understand the structure of these curves in more detail, let us consider for example the sensitivity of LISA to the primary GW signal (solid cyan curve).
At small $f_{\rm PBH}$ the merger rate scales as ${\cal R} \propto f_{\rm PBH}^2$, after incorporating all relevant suppression factors.
We then obtain the bound by imposing $N_{\rm det} \sim 1 \sim {\cal R} \times V$, where $V$ is the integrated cosmological volume within which binaries can be observed. 
For large masses, the SNR is dominated by the contribution near the ISCO frequency, which scales as $f\sim {1}/{m_{\rm PBH}}$,
and below the peak sensitivity the SNR integral behaves as $(f^{-7/3}/f^{-6})^{1/2}$, leading to a strong suppression as a function of mass.\footnote{Here, the $f^{-6}$ scaling in the denominator, as well as the corresponding $f^2$ factor in the following expression reflects the expected scaling of the LISA sensitivity curve, see App.~\ref{app:sens_curves}. These values are detector-specific, but the sensitivity curves can always be parameterized by a red- and blue-tilted power low in the low and high-frequency regimes, respectively.}
However, for masses such that the ISCO frequency lies above LISA's peak sensitivity (roughly $m_{\rm PBH} \lesssim 10^7\,M_\odot$), 
the SNR integral change scaling and goes like $(f^{-7/3}/f^{2})^{1/2}$. 
This modification, together with the effect of the growing cosmological volume at the redshift corresponding to the maximum observable distance, results in an overall bound scaling as 
$f_{\rm PBH}\sim m_{\rm PBH}^{1/2}$. 
At masses below approximately $10^3\,M_\odot$, the duration of the signal in band exceeds the observation time, which reduces the SNR by limiting the number of observable frequency bins. In the small mass limit, primary signals evolve very slowly in frequency and become nearly monochromatic sources, for which the SNR simplifies to $\rho^2 \sim \tilde h^2 (f_{\rm in})/S_h(f_{\rm in})$ \cite{Robson:2018ifk}.
The change of scaling in all constraints around $f_{\rm PBH} \sim 10^{12}$ follows a change in the scaling of the merger rate as a function of $f_{\rm PBH}$.\footnote{
We iterate that this extrapolation to $f_\text{PBH} \gg 1$ is unphysical and we are only interested in a relative comparison between the constraint from inspiral+merger to the one from GW memory: in the low-mass limit, the memory signal can exceed the primary signal. If more sensitive detectors, admittedly quite futuristic, were to become available, all these lines would shift down into the physically meaningful parameter space.}

Similar considerations can be made for the other experiments. As we can see, only the next-generation ground-based detector ET would be able to seriously constrain subsolar sources within our galaxy.
These forecasts agree with the ones previously derived for inspiraling PBH sources
in Ref.~\cite{DeLuca:2021hde,Pujolas:2021yaw}.\footnote{We do not consider extreme mass ratio inspirals, which were shown to extend sensitivities to lighter secondary masses in the binaries, see e.g. \cite{Miller:2024khl}, as we don't expect this to affect the comparison between primary and memory GW signatures.}

Let us now turn to the memory signal, represented by the dashed lines in Fig.~\ref{fig:plots}.
The qualitative behavior of the bounds derived from GW memory resembles that of the primary signal: they are strongest when the effective cutoff frequency $f_{\rm decay}$ approximately aligns with the detector's peak sensitivity and exhibit a secondary enhancement when the detector reach enters the galactic dark matter overdensity. 
Both LISA and ET can detect memory signals in physical scenarios where $f_{\rm PBH} \lesssim 1$, though a realistic assessment must account for existing non-GW constraints on PBH abundance within the relevant mass range.  
In contrast, the memory signal from PBH binaries is likely too weak to be observed by AdLIGO and MWB.  

Notably, for a fixed detector frequency range,  memory becomes increasingly constraining at lower masses, where the number of binaries deep in the inspiral---capable of exciting the low-frequency bands of each detector---declines, while the memory signal from each merger remains detectable. 
In fact, the maximum distance for detecting the memory signal exceeds that of the primary distance already at significantly higher masses than the crossing point shown in Fig.~\ref{fig:plots}. However, since at a given frequency there are fewer merger events (leading to a memory signal) than binaries emitting in the early stages of their inspirals (see Fig.~\ref{fig:dn/df}), the limits on $f_\text{PBH}$ are dominated by the inspiral signals.
This is however specific to the EU formation scenario and, 
as discussed at the end of Sec.~\ref{sec:PBH merger rate}, binaries formed in the late-time universe tend to be much harder, preventing them from exciting low-frequency modes. This makes memory searches for out-of-band signals particularly relevant in such cases.

In this work we have focused on the prediction of single loud memory events, but as for the dominant GW signal, the superposition of unresolved events will generate a stochastic background. It has been shown in Refs.~\cite{Allen:2019hnd, Zhao:2021zlr} that the energy density of such a background decreases faster at low frequencies, $\Omega^{\text{mem}}_{\text{GW}}\propto f$, than for binaries in the inspiral, $\Omega^{\text{ins}}_{\text{GW}}\propto f^{2/3}$. So we expect that for the EU formation scenario the memory background is always subdominant because the cutoff frequency corresponding to the largest binary separation is extremely low (see also \cite{Boybeyi:2024aax}), but we leave a detailed discussion to future work.


\section{Conclusions}\label{sec:conclusions}

The energy flux carried in the emission of a GW burst sources a secondary GW signal, referred to as GW memory, allowing for a complementary search strategy of GW bursts arising for example from mergers of compact objects. This is particularly interesting for high-frequency GW signals, i.e.\ when the primary signal peaks at frequencies above the sensitivity range of a given detector. In this case, the memory signal is relatively universal, with a waveform depending only on the detector properties as well as on the amplitude, phase and sky position of the primary signal. This makes this signal very suitable for an efficient matched filtering search, allowing to simultaneously constrain a broad class of models predicting high-frequency GW bursts. Of course, this same property implies that in the case of a detection, model discrimination or parameter estimation becomes extremely difficult. In this sense, searches for memory signals are similar to searches for extra radiation, often phrased in terms of additional neutrino species $\Delta N_\text{eff}$.

For mergers of black holes, the low-frequency part of the GW signal contains two contributions, sourced at different times in the evolution of the black hole binary: the universal memory signal, emitted predominantly at the merger, and the GWs emitted in the early inspiral stage of the binary. Assuming an optimal matched filtering analysis applied to both signatures and taking into account constraints on the black hole merger rate across cosmic history, we find that the inspiral signal results in a higher signal-to-noise ratio for all current and near-future detectors considered. This can change for mergers which are very far out-of-band, however, for current and near-future detectors  this corresponds to weak signals that would require unrealistically large PBH abundances for a detection.

We conclude that currently, the strongest limits for out-of-band black hole mergers are set by searches for quasi-monochromatic signals sourced in the early inspiral stage. This conclusion applies however only to mergers of compact objects which feature a very coherent low-frequency contribution to the GW signal. Other sources such as PBH encounters~\cite{Zeldovich:1974gvh,1977ApJ...216..610T,Braginsky:1987kwo}, GW from hyperbolic encounters~\cite{Caldarola:2023ipo}, GW bursts from cosmic strings~\cite{Damour:2000wa} and from extreme mass ratio inspirals (EMRIs)~\cite{Amaro-Seoane:2007osp} do not share this property, making a GW memory search a prime candidate for setting limits or for a possible detection. Searching for the memory signal is moreover computationally much less demanding due to the universality of the signal. It also opens up the possibility of coincident searches with detectors sensitive at higher frequencies.

\begin{acknowledgments}
We acknowledge Diego Blas, Ville Vaskonen, Juan Sebastián Valbuena-Bermúdez for their insights during the development of this project.
S.G. thanks the CERN TH department for the hospitality during the completion of this work. The research leading to these results has received funding from the Spanish Ministry of Science and Innovation PID2020-115845GB-I00/AEI/10.13039/501100011033 and PID2023-146686NB-C31 funded by
MICIU/AEI/10.13039/501100011033/ and by FEDER,
UE.
IFAE is partially funded by the CERCA program of the Generalitat de Catalunya. SG has the support of the predoctoral program AGAUR FI SDUR 2022 from the Departament de Recerca i Universitats from Generalitat de Catalunya and the European Social Plus Fund.
This work was supported by the European Research Area
(ERA) via the UNDARK project of the Widening participation and spreading excellence programme (project
number 101159929).
\end{acknowledgments}

\appendix

\section{Gravitational wave memory derivation}
\label{app:memtheory}
As discussed in the main text, the GW memory is the back-reaction of the GW emission on the GW strain itself. In other words, it is the GW component sourced by the energy-momentum tensor carried away by the GWs that changes the quadrupolar and high-order structure of the total system.  This can be viewed as the back-reaction of the high-frequency on the low-frequency component of the metric~\cite{Heisenberg:2023prj}. 

Let's indicate with $h$ the primary GW directly sourced from the energy-momentum tensor of the matter components $T_{\mu\nu}^{\rm mat}$ and with $h_{\rm mem}$ the memory component. The latter solves the back-reaction equations
\begin{equation}\label{eq:EinEqmem}
    G_{\mu\nu}[h_{\rm mem}]= 8\pi T^{\rm gw}_{\mu\nu}, 
\end{equation}
where the sourced term on the right-hand side is the pseudo
energy-momentum tensor of primary GWs which far from the source takes the asymptotic form 
\begin{equation}\label{eq:TGWs}
    T^{\rm gw}_{\mu\nu}= 
    \frac{n_\mu n_\nu}{d^2}\frac{\mathrm{d}E(u,\Omega)}{\mathrm{d}u\mathrm{d}\Omega}=
    \frac{n_\mu n_\nu}{16\pi}\langle \dot{h}_+^2+\dot{h}_\times^2\rangle 
\end{equation}
where $n_\mu= -\nabla_\mu t+ \nabla_\mu d$. Far from the source, the GW flux is purely radially outward and only depends on the retarded time $u=t-d$ and on the angular coordinates with respect to the source $\Omega$.  Note that the triangular brackets in Eq.~\eqref{eq:TGWs} are needed for the proper gauge-invariant definition of the GW energy density (see Sec.~(1.4) of~\cite{maggiore2008gravitational}). By defining, as usual, $\Bar{h}_{\mu\nu}=h_{\mu\nu}-\eta_{\mu\nu}h^\alpha_\alpha/2$, in the harmonic gauge where $\partial^\mu \bar{h}_{\mu\nu}=0$ for the memory as well, Eq.\eqref{eq:EinEqmem} becomes
\begin{equation}
    \Box \Bar{ h}^{\rm mem}_{\mu\nu}=-16\pi T^{\rm gw}_{\mu\nu} .
\end{equation}
This can be solved by the usual Green function method such that at position $\Vec{x}$ it is given by the integral over the whole space where the source is non-zero
\begin{align}\label{eq:memsteps}
    &\Bar{ h}^{\rm mem}_{\mu\nu}(t,\vec{x})=4\int {\rm d}x'^4 T^{\rm gw}_{\mu\nu}(x') \frac{\delta(t-t'-|\Vec{x}-\Vec{x'}|)}{|\Vec{x}-\Vec{x'}|}=
    \notag 
    \\
    &
    = \int {\rm d}r'{\rm d}\Omega' {\rm d}u' n'_\mu n'_\nu\frac{\mathrm{d}E(u',\Omega')}{\mathrm{d}u'\mathrm{d}\Omega'}\frac{\delta(t-t'-d|1-\Hat{n}\cdot \Vec{r'}|)}{d|1-\Hat{n}\cdot \Vec{r'}|}  .
\end{align}
After performing the integration over $r'$ and taking the limit along an outgoing null ray $(d\longrightarrow\infty, u=\text{const})$, one recovers Eq.~\eqref{eq:memoryequation} in the TT gauge which captures the physical modes of the gravitational radiation. 

It is common in the literature to expand
both the main and the memory contribution into spin-weighted spherical harmonics (SWSH) of spin-weight $s=-2$
\begin{equation}\label{eq:SW Memory quantity GR} 
     h(u,r,\Omega)\equiv  h_+-ih_\times=\sum_{l=2}^\infty\sum_{m=-l}^{l}\, h_{lm}(u,r)\,\,_{{-2}}Y_{lm}(\Omega)\,
\end{equation}
where the spatial direction is parametrized by the two angles in the reference frame centered at the source.
 Taking this decomposition the energy flux can be written as  
\begin{equation*}
    |\dot h|^2=\sum_{l_1=2}^\infty\sum_{m_1=-l_1}^{l_1}\sum_{l_2=2}^\infty\sum_{m_2=-l_2}^{l_2} \dot h_{l_1m_1}\dot h^*_{l_2m_2}\,_{{-2}}Y_{l_1m_1}\,_{{-2}}Y^*_{l_2m_2}\,,
\end{equation*}
inserting this into Eq.~\eqref{eq:memoryequation} one can compute the memory from the modes of the primary wave. This can be done with the public \texttt{GWMemory}~\cite{PhysRevD.98.064031} package, which performs the integration over time and over the angular variable of Eq.~\eqref{eq:memoryequation} numerically for some input waveforms decomposed in spherical harmonics. 

The TT projection factor in Eq.~\eqref{eq:memoryequation} can be also written into spherical harmonics, as shown in paragraph IV (C) of Ref.~\cite{Heisenberg:2023prj}. With that the memory modes can be computed explicitly from \cite{Favata:2008yd,Zosso:2024xgy} 
\begin{equation}
    h_{\ell m}^{\rm mem}=d\sum_{\ell',\ell''\geq 2}\, \sum_{m',m''} \Gamma^{l'm'm''l''}_{lm} \int_{-\infty}^u \mathrm{d}u'\langle\dot{h}^{\ell' m'} \dot{h}^{*\ell'' m''}\rangle,
\label{eq:memorymodes}
\end{equation}
with
\begin{align*}
    \Gamma^{l'm'm''l''}_{lm}&\equiv (-1)^{m+m''}\sqrt{\frac{(2l'+1)(2l''+1)(2l+1)}{4\pi}}\\
    &\times\sqrt{\frac{(l-2)!}{(l+2)!}}
    \begin{pmatrix}
        l' & l'' & l\\
         m' & -m'' & -m
    \end{pmatrix}
    \begin{pmatrix}
        l' & l'' & l\\
        2 & -2 & 0
    \end{pmatrix},
\end{align*}
where the big parenthesis represents the Wigner $3-j$ symbols, which in this case are only non-zero if $m=m'-m''$ and $|l'-l''|\leq l\leq l'+l''$. From the selection rule above it becomes evident that for a quasi-circular and non-precessing signal, for which the leading GW modes are the $h_{(2,\pm2)}$, it is easy to see that the memory is predominantly found in the $m=0$ modes with $2\leq l \leq 4$. Therefore, one finds that the leading memory contributions are
\begin{equation*}
    h^{\rm mem}_{(2,0)}
    =\frac{d}{7}\sqrt{\frac{5}{6\pi}}
    \int_{-\infty}^{u}
    \d u' \langle |\dot{h}_{(2,2)}|^2\rangle 
    = 60\sqrt{3} h^{\rm mem}_{(4,0)} \,.
\end{equation*}
Using Eq.~\eqref{eq:SW Memory quantity GR} and the fact that $h_+=Re[h]$ and $h_\times=Im[h]$, in this coordinate system the memory just affects one polarization, which is conventionally chosen to be the $+$ polarization in the source frame with the $z$-axis pointing in the direction of the observer, and the binary plane intersecting the $x$-$y$-plane at an inclination angle $\iota$ (see e.g.~\cite{favata_nonlinear_2009}),
\begin{equation}\label{eq:crossmem}
    h^{\rm mem}_+=\frac{d}{192\pi}\sin^2{\iota}(17+\cos^2{\iota})\int_{-\infty}^{u}
    \d u' 
    \langle|\dot{h}_{(2,2)}|^2\rangle
\end{equation}
and $h^{\rm mem}_\times=0$. 

In the SWSH decomposition, the modes take the form $h_{\ell,m}=A_{\ell,m} e^{-i\phi_{\ell,m}}$ where $A_{\ell,m}$ is the mode's amplitude and $\phi_{\ell,m}$ is the phase, thus the integrand in Eq.~\eqref{eq:crossmem} is not oscillating explaining why the leading memory contribution is not periodic and its time dependence is given by $A_{\ell,m}$ that, in the inspiral, evolves over the radiation reaction timescale. Because of this feature, the average in Eq.~\eqref{eq:crossmem} can drop off, but it will have a non-negligible effect for eccentric or precessing waveforms.

\section{Memory signals at interferometers}
\label{app:response}
This appendix derives the signal templates for GW memory at LIGO and LISA in the out-of-band limit $f \ll \Delta \tau^{-1}$. For simplicity we set $t_m = 0$. Our starting point is the memory signal,
\begin{align}
 h_{ab}(t, \vec x)  =
 & = e^+_{ab}(\hat k) \int_{- \infty}^{\infty} df \, \tilde h(f) \, e^{2 \pi i f (t - \hat k \vec x)}\,,
\end{align}
with the Fourier coefficients given in Eq.~\eqref{eq:FTmem} and the polarisation tensor for the plus polarization defined as
\begin{align}
 e_{ab}^+ = u_a u_b - v_a v_b \,,
\end{align}
with the unit vectors ${\bm u}$ and ${\bm v}$,
\begin{align}
 \quad {\bm u} = (- \sin \phi, - \cos \phi, 0) \,, \quad {\bm v} = \hat{\bm k} \times {\bm u} \,,
\end{align}
forming an orthonormal system with respect to the gravitational wave vector
\begin{align}
 \hat{\bm k} = ( \sin \theta \cos \phi, \sin \theta \sin \phi, \cos \theta ) \,.
\end{align}

 For the detector response, we follow Ref.~\cite{Romano:2016dpx}. The time delay induced by a GW during a return trip of light along an interferometer arm of length $L$ and orientation $\hat{\bm \ell}$ is
\begin{align}
 \Delta T(t, \hat{\bm \ell}, {\bm x}_1) = \frac{\hat \ell^a \hat \ell^b}{2}  e_{ab}^+ \int_{-\infty}^\infty \! df \, {\cal T}(\hat{\bm k}\cdot\hat{\bm \ell}, f) \, \tilde h(f) \, e^{2 \pi i f (t - \hat{\bm k} \cdot \bm{x}_1)}.
 \label{eq:DeltaT}
\end{align}
Since we are considering only a single detector, we can set the location of the beamsplitter ${\bm x}_1 = 0$. The frequency dependent part of the response function is given by
\begin{align}
 {\cal T}(\hat{\bm k}\cdot\hat{\bm \ell}, f) = L e^{- 2 \pi i f L} & \left[ e^{- i \pi f L (1 + \hat{\bm k}\cdot\hat{\bm \ell})} \text{sinc}[ \pi f L (1 - \hat{\bm k}\cdot\hat{\bm \ell})] \right. \nonumber \\
 &  \left. +  e^{i \pi f L (1 - \hat{\bm k}\cdot\hat{\bm \ell})} \text{sinc}[ \pi f L (1 + \hat{\bm k}\cdot\hat{\bm \ell})] \right]
\end{align}
and we note ${\cal T}(\hat{\bm k}\cdot\hat{\bm \ell}, - f) = {\cal T}^*(\hat{\bm k}\cdot\hat{\bm \ell}, f)$. With $\tilde h(-f) = \tilde h^*(f)$ this ensures that $\Delta T$ is real. In the low-frequency limit we obtain  ${\cal T}(\hat{\bm k}\cdot\hat{\bm \ell}, 0)/L = 2$.

The final signal in the detector is given by the difference in the time delays in the two arms,
\begin{align}
 s(t) & = (\Delta T(t, \hat{\bm \ell}_{13}) - \Delta T(t, \hat{\bm \ell}_{12}))/(2 L)\,, \nonumber \\
 & = \int_{- \infty}^\infty df R(\bm k) \tilde h(f) e^{2 \pi i f t} \,.
 \label{eq:s_interferometer}
\end{align}
Here we have introduced the response function
\begin{align}
 R(\bm k) = \frac{e_{ab}^+}{4 L} \left[ \hat{\bm \ell}_{13}^a \hat{\bm \ell}_{13}^b {\cal T}(\hat{\bm k}\cdot\hat{\bm \ell}_{13}, f) - \hat{\bm \ell}_{12}^a \hat{\bm \ell}_{12}^b {\cal T}(\hat{\bm k}\cdot\hat{\bm \ell}_{12}, f) \right].
\end{align}
In the low-frequency limit, $f \ll 1/L$, this simplifies to
\begin{align}
 R(\bm k) \rightarrow \frac{e_{ab}^+}{2} \left[ \hat{\bm \ell}_{13}^a \hat{\bm \ell}_{13}^b - \hat{\bm \ell}_{12}^a \hat{\bm \ell}_{12}^b \right]\,.
\end{align}
For an equilateral triangular configuration, such as LISA, placed in the $x$-$y$ plane, $R(\hat e_z) = 3/4$. For LIGO, $R(\hat e_z) = - 1$.

Fig.~\ref{fig:bandpassed memory} in the main text shows some examples of the resulting signal in LIGO. For LIGO, $f_{\rm min}^{\rm det} = 10$~Hz is set by a bandpass filter used in the data processing to remove the frequency range dominated by seismic noise, while $f_{\rm max}^{\rm det} = 8192$~Hz is set by the Nyquist frequency. Since most of LIGO's frequency range is in the low frequency limit, $f L \lesssim 1$, the dependence on the sky position is only mild. It is worth noting that for LIGO, $f_{\rm min}^{\rm det}$, $f_{\rm max}^{\rm det}$ and $f_\text{arm} \sim 1/L = 250$~Hz are all rather close together, resulting in the waveform depicted in Fig.~\ref{fig:bandpassed memory}.

\begin{figure}
\centering
 \includegraphics[width = 0.48 \textwidth]{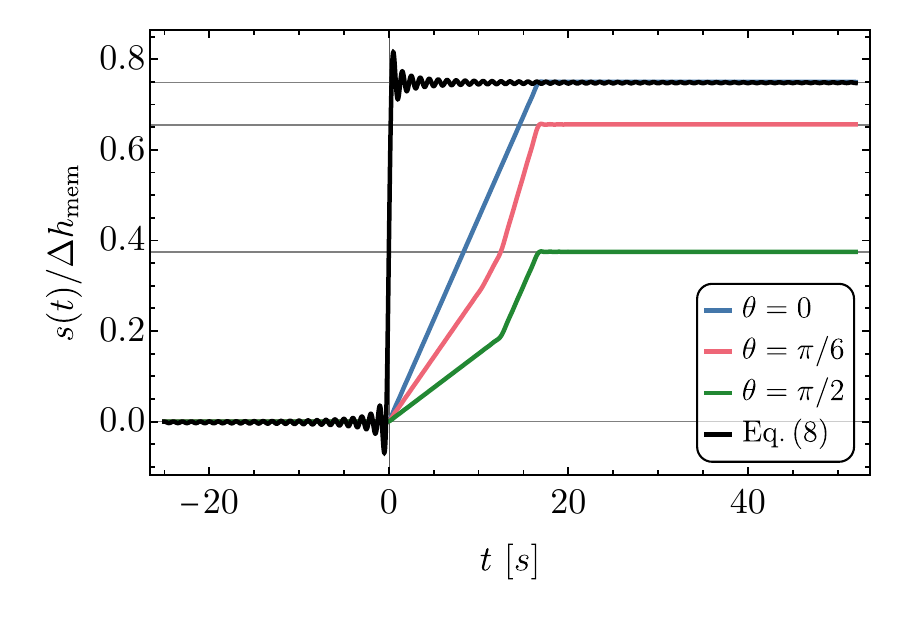} \hfill
 \includegraphics[width = 0.48 \textwidth]{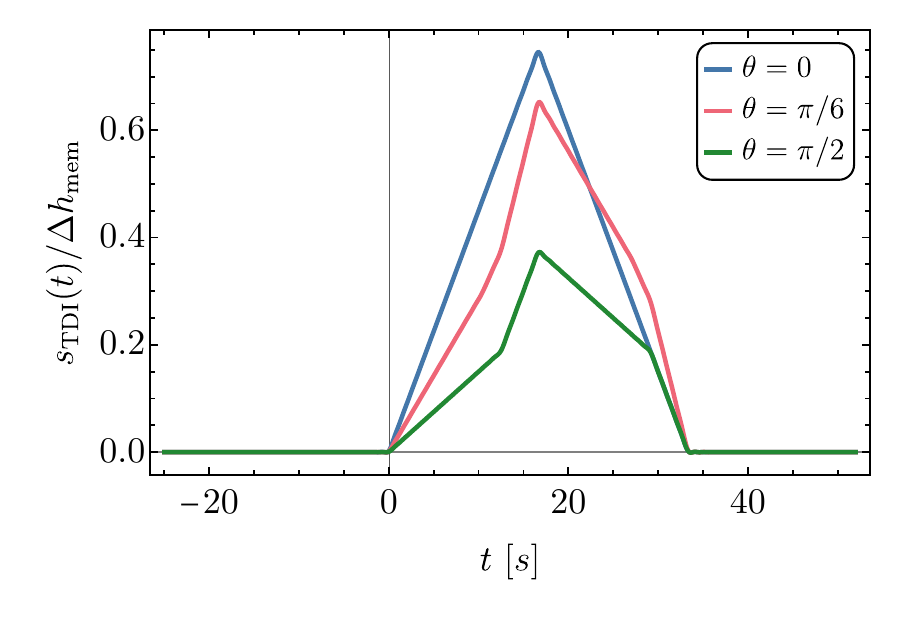}
 \caption{GW signal in LISA for $\theta = \{0, \pi/6, \pi/2\}$ (blue, yellow, green) and $\phi = 0$, compared to the approximate expression Eq.~\eqref{eq:s_approx} (black, gray). Here $\hat{\bm \ell}_{12} = \hat e_x$, $\hat{\bm \ell}_{13} = (1/2, \sqrt{3/4}, 0)$. The upper panel is for a simple equilateral interferometer, the lower panel uses TDI variables, $s_\text{TDI}(t) = s(t) - s(t - 2 L)$.}
 \label{fig:LISA}
\end{figure}

Turning to the case of LISA, we must take a further step to derive the signal in the TDI (time delayed interferometry) variables,
\begin{align}
 s_\text{TDI} = s(t) - s(t - 2L) \,.
\end{align}
This is necessary to cancel the dominant laser noise arising, which would otherwise contribute a noise source which is by orders of magnitude larger. Fig.~\ref{fig:LISA} shows the resulting signal both for a simple equilateral interferometer and after implementing TDI variables. Notably, and as discussed also in Ref.~\cite{Ghosh:2023rbe}, the use of TDI variables by construction cancels all features in the signal that come with a frequency much smaller than $1/L$, so that we are only left with the characteristic memory signal associated with the light travel time in the interferometer arms. Moreover,
 LISA features a very broad frequency range, set by the inverse of the (uninterrupted) observation time and the Nyquist frequency, $f \in [10^{-6},1]$~Hz, while $f_\text{arm} \sim 1/L = 0.1$~Hz. The large separation between $f_{\rm min}^{\rm det}$ and $f_\text{arm}$ explains why the $f_{\rm min}^{\rm det}$ oscillations are not visible in the upper panel of Fig.~\ref{fig:LISA}. Moreover, we note that since LISA is not operating in the low-frequency limit, the dependence of the waveform on the sky position is more pronounced.

\section{Power spectral density sensitivity curves}
\label{app:sens_curves}

\begin{figure}[t!]
    \centering
    \includegraphics[width = \linewidth]{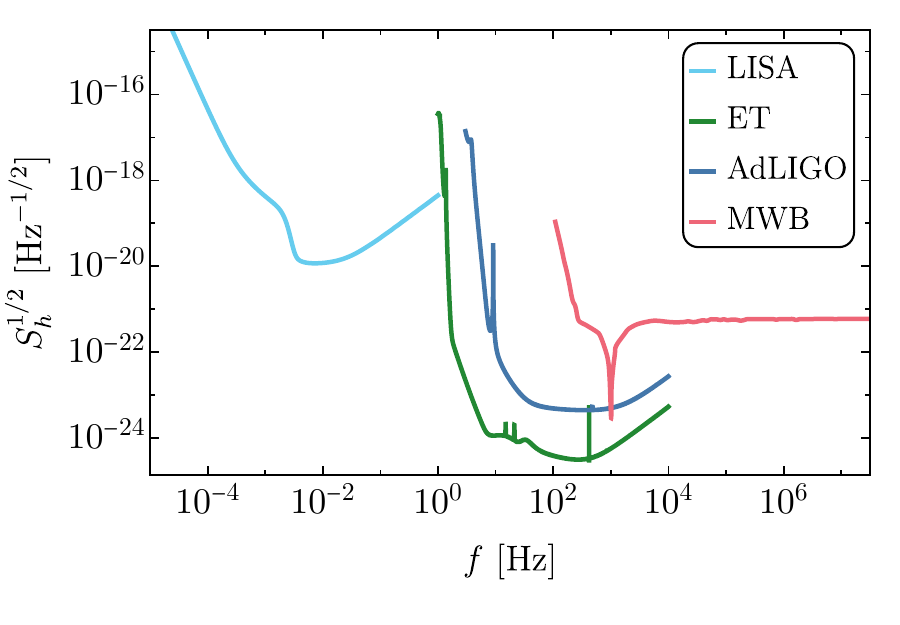}
    \caption{ 
    Power spectral density for the various GW experiments considered in this work. 
    From left to right: 
    LISA (cyan), 
    ET (green), 
    Advanced LIGO (blue), 
    and Magnetic Weber Bars (red). 
    } 
    \label{fig:psd}
\end{figure}

The noise power spectral densities adopted in this work are shown in Fig.~\ref{fig:psd}.
In particular, we consider:
\begin{itemize}[leftmargin=2.5mm]
    \item Advanced LIGO: we adopt the power spectral density $P_n(f)$ of the design sensitivity~\cite{Barsotti:18}.
    In this case, to get the noise-equivalent strain power
spectral density of the detector $S_n = P_n / R^2$, we compute the sky/polarization average of the response in the low frequency limit (i.e. the antenna patterns), which gives $ R^2 \to  \langle F_{+}^2\rangle =\langle F_{\times}^2\rangle=1/5$;
    \item Einstein Telescope: we take the power spectral density $P_n(f)$ to be the ET-D curve from Ref.~\cite{Hild:2010id}. In this case,  $  R^2 \to \langle F_{+}^2\rangle =\langle F_{\times}^2\rangle=3/20 $ for a triangular configuration;
    \item Magnetic Weber Bars: we follow the proposal of Ref.~\cite{Domcke:2024mfu} that showed a superconducting magnet, operated in persistent mode, acts as a resonant mass detector. We adopt the sensitivity estimated for a magnet of the size proposed for the DMRadio-GUT
    \cite{DMRadio:2022jfv} detector. 
    \item LISA:
    We consider the LISA noise-equivalent strain power
spectral density $S_n$ of Ref.~\cite{Robson:2018ifk}, which is expressed as an analytic fit for the detector noise. This consists of two parts: the instrumental noise and the confusion noise produced by unresolved galactic binaries, i.e. 
\begin{equation}
S_{n}(f)=S^\tn{Ins}_{n}(f)+S^\tn{WDN}_{n}(f) \ ,
\end{equation}
where 
\begin{align}
S^\tn{Ins}_{n}(f)&=A_1\left(P_\tn{OMS}+2[1+\cos^2(f/f_\star)]
\frac{P_\tn{acc}}{(2\pi f)^4}\right)
\nonumber\\
&\times \left(1+\frac{6}{10}\frac{f^2}{f_\star^2}\right)\ ,
\end{align}
$A_1={10}/{3L^2}$, $L=2.5$Gm, $f_\star=19.09$mHz, while 
\begin{align}
P_\tn{OMS}&=(1.5\times 10^{-11}\tn{m})^2\left[1+\left(\frac{2\tn{mHz}}{f}\right)^4\right]\ \tn{Hz}^{-1}\nonumber\ ,\\ 
P_\tn{ACC}&=(3\times 10^{-15}\tn{m s}^{-2})^2\left[1+\left(\frac{0.4\tn{mHz}}{f}\right)^2\right]
\nonumber \\
&\times \left[1+\left(\frac{f}{8\tn{mHz}}\right)^4\right]\ \tn{Hz}^{-1} .
\end{align}
For the white dwarf contribution, we use
\begin{equation}
S^\tn{WDN}_{n}=A_2 f^{-7/3}e^{-f^\alpha+\beta f\sin(\kappa f)}[1+\tanh(\gamma(f_k-f))]\ ,
\end{equation}
with the amplitude $A_2=9\times 10^{-45} \tn{Hz}^{-1}$, and the coefficients 
$(\alpha,\beta,\kappa,\gamma,f_k)=(0.171,292,1020,1680,0.00215)$.
\end{itemize}

\bibliographystyle{utphys}
\bibliography{refs}

\providecommand{\href}[2]{#2}\begingroup\raggedright\begin{thebibliography}{100}

\bibitem{LIGOScientific:2016aoc}
{\bfseries LIGO Scientific, Virgo} Collaboration, B.~P. Abbott {\em et~al.}, ``{Observation of Gravitational Waves from a Binary Black Hole Merger},'' \href{https://dx.doi.org/10.1103/PhysRevLett.116.061102}{{\em Phys. Rev. Lett.} {\bfseries 116} no.~6, (2016) 061102}, \href{https://arxiv.org/abs/1602.03837}{{\ttfamily arXiv:1602.03837 [gr-qc]}}.

\bibitem{KAGRA:2021vkt}
{\bfseries KAGRA, VIRGO, LIGO Scientific} Collaboration, R.~Abbott {\em et~al.}, ``{GWTC-3: Compact Binary Coalescences Observed by LIGO and Virgo during the Second Part of the Third Observing Run},'' \href{https://dx.doi.org/10.1103/PhysRevX.13.041039}{{\em Phys. Rev. X} {\bfseries 13} no.~4, (2023) 041039}, \href{https://arxiv.org/abs/2111.03606}{{\ttfamily arXiv:2111.03606 [gr-qc]}}.

\bibitem{Xu:2023wog}
H.~Xu {\em et~al.}, ``{Searching for the Nano-Hertz Stochastic Gravitational Wave Background with the Chinese Pulsar Timing Array Data Release I},'' \href{https://dx.doi.org/10.1088/1674-4527/acdfa5}{{\em Res. Astron. Astrophys.} {\bfseries 23} no.~7, (2023) 075024}, \href{https://arxiv.org/abs/2306.16216}{{\ttfamily arXiv:2306.16216 [astro-ph.HE]}}.

\bibitem{EPTA:2023fyk}
{\bfseries EPTA, InPTA:} Collaboration, J.~Antoniadis {\em et~al.}, ``{The second data release from the European Pulsar Timing Array - III. Search for gravitational wave signals},'' \href{https://dx.doi.org/10.1051/0004-6361/202346844}{{\em Astron. Astrophys.} {\bfseries 678} (2023) A50}, \href{https://arxiv.org/abs/2306.16214}{{\ttfamily arXiv:2306.16214 [astro-ph.HE]}}.

\bibitem{NANOGrav:2023gor}
{\bfseries NANOGrav} Collaboration, G.~Agazie {\em et~al.}, ``{The NANOGrav 15 yr Data Set: Evidence for a Gravitational-wave Background},'' \href{https://dx.doi.org/10.3847/2041-8213/acdac6}{{\em Astrophys. J. Lett.} {\bfseries 951} no.~1, (2023) L8}, \href{https://arxiv.org/abs/2306.16213}{{\ttfamily arXiv:2306.16213 [astro-ph.HE]}}.

\bibitem{Reardon:2023gzh}
D.~J. Reardon {\em et~al.}, ``{Search for an Isotropic Gravitational-wave Background with the Parkes Pulsar Timing Array},'' \href{https://dx.doi.org/10.3847/2041-8213/acdd02}{{\em Astrophys. J. Lett.} {\bfseries 951} no.~1, (2023) L6}, \href{https://arxiv.org/abs/2306.16215}{{\ttfamily arXiv:2306.16215 [astro-ph.HE]}}.

\bibitem{Miles:2024rjc}
M.~T. Miles {\em et~al.}, ``{The MeerKAT Pulsar Timing Array: The $4.5$-year data release and the noise and stochastic signals of the millisecond pulsar population},'' \href{https://dx.doi.org/10.1093/mnras/stae2572}{{\em Mon. Not. Roy. Astron. Soc.} {\bfseries 536} no.~2, (2025) 1467--1488}, \href{https://arxiv.org/abs/2412.01148}{{\ttfamily arXiv:2412.01148 [astro-ph.HE]}}.

\bibitem{EPTA:2023xxk}
{\bfseries EPTA, InPTA} Collaboration, J.~Antoniadis {\em et~al.}, ``{The second data release from the European Pulsar Timing Array - IV. Implications for massive black holes, dark matter, and the early Universe},'' \href{https://dx.doi.org/10.1051/0004-6361/202347433}{{\em Astron. Astrophys.} {\bfseries 685} (2024) A94}, \href{https://arxiv.org/abs/2306.16227}{{\ttfamily arXiv:2306.16227 [astro-ph.CO]}}.

\bibitem{Zeldovich:1967lct}
Y.~B. Zel'dovich and I.~D. Novikov, ``{The Hypothesis of Cores Retarded during Expansion and the Hot Cosmological Model},'' {\em Sov. Astron.} {\bfseries 10} (1967) 602.

\bibitem{Hawking:1971ei}
S.~Hawking, ``{Gravitationally collapsed objects of very low mass},'' \href{https://dx.doi.org/10.1093/mnras/152.1.75}{{\em Mon. Not. Roy. Astron. Soc.} {\bfseries 152} (1971) 75}.

\bibitem{Carr:1974nx}
B.~J. Carr and S.~W. Hawking, ``{Black holes in the early Universe},'' \href{https://dx.doi.org/10.1093/mnras/168.2.399}{{\em Mon. Not. Roy. Astron. Soc.} {\bfseries 168} (1974) 399--415}.

\bibitem{Carr:1975qj}
B.~J. Carr, ``{The Primordial black hole mass spectrum},'' \href{https://dx.doi.org/10.1086/153853}{{\em Astrophys. J.} {\bfseries 201} (1975) 1--19}.

\bibitem{Byrnes:2025tji}
C.~Byrnes, G.~Franciolini, T.~Harada, P.~Pani, and M.~Sasaki, eds., {\em {Primordial Black Holes}}.
\newblock Springer Series in Astrophysics and Cosmology. Springer, 3, 2025.

\bibitem{Carr:2020gox}
B.~Carr, K.~Kohri, Y.~Sendouda, and J.~Yokoyama, ``{Constraints on primordial black holes},'' \href{https://dx.doi.org/10.1088/1361-6633/ac1e31}{{\em Rept. Prog. Phys.} {\bfseries 84} no.~11, (2021) 116902}, \href{https://arxiv.org/abs/2002.12778}{{\ttfamily arXiv:2002.12778 [astro-ph.CO]}}.

\bibitem{Bird:2016dcv}
S.~Bird, I.~Cholis, J.~B. Mu\~noz, Y.~Ali-Ha\"\i{}moud, M.~Kamionkowski, E.~D. Kovetz, A.~Raccanelli, and A.~G. Riess, ``{Did LIGO detect dark matter?},'' \href{https://dx.doi.org/10.1103/PhysRevLett.116.201301}{{\em Phys. Rev. Lett.} {\bfseries 116} no.~20, (2016) 201301}, \href{https://arxiv.org/abs/1603.00464}{{\ttfamily arXiv:1603.00464 [astro-ph.CO]}}.

\bibitem{Clesse:2016vqa}
S.~Clesse and J.~Garc\'\i{}a-Bellido, ``{The clustering of massive Primordial Black Holes as Dark Matter: measuring their mass distribution with Advanced LIGO},'' \href{https://dx.doi.org/10.1016/j.dark.2016.10.002}{{\em Phys. Dark Univ.} {\bfseries 15} (2017) 142--147}, \href{https://arxiv.org/abs/1603.05234}{{\ttfamily arXiv:1603.05234 [astro-ph.CO]}}.

\bibitem{Sasaki:2016jop}
M.~Sasaki, T.~Suyama, T.~Tanaka, and S.~Yokoyama, ``{Primordial Black Hole Scenario for the Gravitational-Wave Event GW150914},'' \href{https://dx.doi.org/10.1103/PhysRevLett.117.061101}{{\em Phys. Rev. Lett.} {\bfseries 117} no.~6, (2016) 061101}, \href{https://arxiv.org/abs/1603.08338}{{\ttfamily arXiv:1603.08338 [astro-ph.CO]}}. [Erratum: Phys.Rev.Lett. 121, 059901 (2018)].

\bibitem{Eroshenko:2016hmn}
Y.~N. Eroshenko, ``{Gravitational waves from primordial black holes collisions in binary systems},'' \href{https://dx.doi.org/10.1088/1742-6596/1051/1/012010}{{\em J. Phys. Conf. Ser.} {\bfseries 1051} no.~1, (2018) 012010}, \href{https://arxiv.org/abs/1604.04932}{{\ttfamily arXiv:1604.04932 [astro-ph.CO]}}.

\bibitem{Wang:2016ana}
S.~Wang, Y.-F. Wang, Q.-G. Huang, and T.~G.~F. Li, ``{Constraints on the Primordial Black Hole Abundance from the First Advanced LIGO Observation Run Using the Stochastic Gravitational-Wave Background},'' \href{https://dx.doi.org/10.1103/PhysRevLett.120.191102}{{\em Phys. Rev. Lett.} {\bfseries 120} no.~19, (2018) 191102}, \href{https://arxiv.org/abs/1610.08725}{{\ttfamily arXiv:1610.08725 [astro-ph.CO]}}.

\bibitem{Ali-Haimoud:2017rtz}
Y.~Ali-Ha\"\i{}moud, E.~D. Kovetz, and M.~Kamionkowski, ``{Merger rate of primordial black-hole binaries},'' \href{https://dx.doi.org/10.1103/PhysRevD.96.123523}{{\em Phys. Rev. D} {\bfseries 96} no.~12, (2017) 123523}, \href{https://arxiv.org/abs/1709.06576}{{\ttfamily arXiv:1709.06576 [astro-ph.CO]}}.

\bibitem{Chen:2018czv}
Z.-C. Chen and Q.-G. Huang, ``{Merger Rate Distribution of Primordial-Black-Hole Binaries},'' \href{https://dx.doi.org/10.3847/1538-4357/aad6e2}{{\em Astrophys. J.} {\bfseries 864} no.~1, (2018) 61}, \href{https://arxiv.org/abs/1801.10327}{{\ttfamily arXiv:1801.10327 [astro-ph.CO]}}.

\bibitem{Raidal:2018bbj}
M.~Raidal, C.~Spethmann, V.~Vaskonen, and H.~Veerm\"ae, ``{Formation and Evolution of Primordial Black Hole Binaries in the Early Universe},'' \href{https://dx.doi.org/10.1088/1475-7516/2019/02/018}{{\em JCAP} {\bfseries 02} (2019) 018}, \href{https://arxiv.org/abs/1812.01930}{{\ttfamily arXiv:1812.01930 [astro-ph.CO]}}.

\bibitem{Franciolini:2021tla}
G.~Franciolini, V.~Baibhav, V.~De~Luca, K.~K.~Y. Ng, K.~W.~K. Wong, E.~Berti, P.~Pani, A.~Riotto, and S.~Vitale, ``{Searching for a subpopulation of primordial black holes in LIGO-Virgo gravitational-wave data},'' \href{https://dx.doi.org/10.1103/PhysRevD.105.083526}{{\em Phys. Rev. D} {\bfseries 105} no.~8, (2022) 083526}, \href{https://arxiv.org/abs/2105.03349}{{\ttfamily arXiv:2105.03349 [gr-qc]}}.

\bibitem{Franciolini:2021nvv}
G.~Franciolini, \href{https://dx.doi.org/10.13097/archive-ouverte/unige:156136}{{\em {Primordial Black Holes: from Theory to Gravitational Wave Observations}}}.
\newblock PhD thesis, Geneva U., Dept. Theor. Phys., 2021.
\newblock \href{https://arxiv.org/abs/2110.06815}{{\ttfamily arXiv:2110.06815 [astro-ph.CO]}}.

\bibitem{Afroz:2024fzp}
S.~Afroz and S.~Mukherjee, ``{Phase Space of Binary Black Holes from Gravitational Wave Observations to Unveil its Formation History},'' \href{https://arxiv.org/abs/2411.07304}{{\ttfamily arXiv:2411.07304 [astro-ph.HE]}}.

\bibitem{Crescimbeni:2024qrq}
F.~Crescimbeni, G.~Franciolini, P.~Pani, and M.~Vaglio, ``{Cosmology and nuclear-physics implications of a subsolar gravitational-wave event},'' \href{https://arxiv.org/abs/2408.14287}{{\ttfamily arXiv:2408.14287 [astro-ph.HE]}}.

\bibitem{Franciolini:2022htd}
G.~Franciolini, A.~Maharana, and F.~Muia, ``{Hunt for light primordial black hole dark matter with ultrahigh-frequency gravitational waves},'' \href{https://dx.doi.org/10.1103/PhysRevD.106.103520}{{\em Phys. Rev. D} {\bfseries 106} no.~10, (2022) 103520}, \href{https://arxiv.org/abs/2205.02153}{{\ttfamily arXiv:2205.02153 [astro-ph.CO]}}.

\bibitem{Aggarwal:2025noe}
N.~Aggarwal {\em et~al.}, ``{Challenges and Opportunities of Gravitational Wave Searches above 10 kHz},'' \href{https://arxiv.org/abs/2501.11723}{{\ttfamily arXiv:2501.11723 [gr-qc]}}.

\bibitem{Franciolini:2023opt}
G.~Franciolini, F.~Iacovelli, M.~Mancarella, M.~Maggiore, P.~Pani, and A.~Riotto, ``{Searching for primordial black holes with the Einstein Telescope: Impact of design and systematics},'' \href{https://dx.doi.org/10.1103/PhysRevD.108.043506}{{\em Phys. Rev. D} {\bfseries 108} no.~4, (2023) 043506}, \href{https://arxiv.org/abs/2304.03160}{{\ttfamily arXiv:2304.03160 [gr-qc]}}.

\bibitem{Abac:2025saz}
A.~Abac {\em et~al.}, ``{The Science of the Einstein Telescope},'' \href{https://arxiv.org/abs/2503.12263}{{\ttfamily arXiv:2503.12263 [gr-qc]}}.

\bibitem{LISACosmologyWorkingGroup:2023njw}
{\bfseries LISA Cosmology Working Group} Collaboration, E.~Bagui {\em et~al.}, ``{Primordial black holes and their gravitational-wave signatures},'' \href{https://dx.doi.org/10.1007/s41114-024-00053-w}{{\em Living Rev. Rel.} {\bfseries 28} no.~1, (2025) 1}, \href{https://arxiv.org/abs/2310.19857}{{\ttfamily arXiv:2310.19857 [astro-ph.CO]}}.

\bibitem{LISA:2024hlh}
{\bfseries LISA} Collaboration, M.~Colpi {\em et~al.}, ``{LISA Definition Study Report},'' \href{https://arxiv.org/abs/2402.07571}{{\ttfamily arXiv:2402.07571 [astro-ph.CO]}}.

\bibitem{Christodoulou:1991cr}
D.~Christodoulou, ``{Nonlinear nature of gravitation and gravitational wave experiments},'' \href{https://dx.doi.org/10.1103/PhysRevLett.67.1486}{{\em Phys. Rev. Lett.} {\bfseries 67} (1991) 1486--1489}.

\bibitem{Blanchet:1992br}
L.~Blanchet and T.~Damour, ``{Hereditary effects in gravitational radiation},'' \href{https://dx.doi.org/10.1103/PhysRevD.46.4304}{{\em Phys. Rev. D} {\bfseries 46} (1992) 4304--4319}.

\bibitem{Thorne:1992sdb}
K.~S. Thorne, ``{Gravitational-wave bursts with memory: The Christodoulou effect},'' \href{https://dx.doi.org/10.1103/PhysRevD.45.520}{{\em Phys. Rev. D} {\bfseries 45} no.~2, (1992) 520--524}.

\bibitem{Mitman:2024uss}
K.~Mitman {\em et~al.}, ``{A review of gravitational memory and BMS frame fixing in numerical relativity},'' \href{https://dx.doi.org/10.1088/1361-6382/ad83c2}{{\em Class. Quant. Grav.} {\bfseries 41} no.~22, (2024) 223001}, \href{https://arxiv.org/abs/2405.08868}{{\ttfamily arXiv:2405.08868 [gr-qc]}}.

\bibitem{McNeill:2017uvq}
L.~O. McNeill, E.~Thrane, and P.~D. Lasky, ``{Detecting Gravitational Wave Memory without Parent Signals},'' \href{https://dx.doi.org/10.1103/PhysRevLett.118.181103}{{\em Phys. Rev. Lett.} {\bfseries 118} no.~18, (2017) 181103}, \href{https://arxiv.org/abs/1702.01759}{{\ttfamily arXiv:1702.01759 [astro-ph.IM]}}.

\bibitem{maggiore2008gravitational}
M.~Maggiore, \href{https://dx.doi.org/10.1093/acprof:oso/9780198570745.001.0001}{{\em Gravitational Waves: Volume 1: Theory and Experiments}}.
\newblock Oxford University Press, 10, 2007.

\bibitem{Ghosh:2023rbe}
S.~Ghosh, A.~Weaver, J.~Sanjuan, P.~Fulda, and G.~Mueller, ``{Detection of the gravitational memory effect in LISA using triggers from ground-based detectors},'' \href{https://dx.doi.org/10.1103/PhysRevD.107.084051}{{\em Phys. Rev. D} {\bfseries 107} no.~8, (2023) 084051}.

\bibitem{Ebersold:2020zah}
M.~Ebersold and S.~Tiwari, ``{Search for nonlinear memory from subsolar mass compact binary mergers},'' \href{https://dx.doi.org/10.1103/PhysRevD.101.104041}{{\em Phys. Rev. D} {\bfseries 101} no.~10, (2020) 104041}, \href{https://arxiv.org/abs/2005.03306}{{\ttfamily arXiv:2005.03306 [gr-qc]}}.

\bibitem{Zeldovich:1974gvh}
Y.~B. Zel'dovich and A.~G. Polnarev, ``{Radiation of gravitational waves by a cluster of superdense stars},'' {\em Sov. Astron.} {\bfseries 18} (1974) 17.

\bibitem{Braginsky:1985vlg}
V.~B. Braginsky and L.~P. Grishchuk, ``{Kinematic Resonance and Memory Effect in Free Mass Gravitational Antennas},'' {\em Sov. Phys. JETP} {\bfseries 62} (1985) 427--430.

\bibitem{Braginsky:1987}
V.~B. {Braginsky} and K.~S. {Thorne}, ``{Gravitational-wave bursts with memory and experimental prospects},'' \href{https://dx.doi.org/10.1038/327123a0}{{\em \nat} {\bfseries 327} no.~6118, (1987) 123--125}.

\bibitem{Vartanyan_2020}
D.~Vartanyan and A.~Burrows, ``Gravitational waves from neutrino emission asymmetries in core-collapse supernovae,'' \href{https://dx.doi.org/10.3847/1538-4357/abafac}{{\em The Astrophysical Journal} {\bfseries 901} no.~2, (2020) 108}. \url{https://dx.doi.org/10.3847/1538-4357/abafac}.

\bibitem{Flanagan:2018yzh}
E.~E. Flanagan, A.~M. Grant, A.~I. Harte, and D.~A. Nichols, ``{Persistent gravitational wave observables: general framework},'' \href{https://dx.doi.org/10.1103/PhysRevD.99.084044}{{\em Phys. Rev. D} {\bfseries 99} no.~8, (2019) 084044}.

\bibitem{Grant:2021hga}
A.~M. Grant and D.~A. Nichols, ``{Persistent gravitational wave observables: Curve deviation in asymptotically flat spacetimes},'' \href{https://dx.doi.org/10.1103/PhysRevD.105.024056}{{\em Phys. Rev. D} {\bfseries 105} no.~2, (2022) 024056}. [Erratum: Phys.Rev.D 107, 109902 (2023)].

\bibitem{Grant:2022bla}
A.~M. Grant and D.~A. Nichols, ``{Outlook for detecting the gravitational-wave displacement and spin memory effects with current and future gravitational-wave detectors},'' \href{https://dx.doi.org/10.1103/PhysRevD.107.064056}{{\em Phys. Rev. D} {\bfseries 107} no.~6, (2023) 064056}, \href{https://arxiv.org/abs/2210.16266}{{\ttfamily arXiv:2210.16266 [gr-qc]}}. [Erratum: Phys.Rev.D 108, 029901 (2023)].

\bibitem{Siddhant:2024nft}
S.~Siddhant, A.~M. Grant, and D.~A. Nichols, ``{Higher memory effects and the post-Newtonian calculation of their gravitational-wave signals},''.

\bibitem{Pasterski:2015tva}
S.~Pasterski, A.~Strominger, and A.~Zhiboedov, ``{New Gravitational Memories},'' \href{https://dx.doi.org/10.1007/JHEP12(2016)053}{{\em JHEP} {\bfseries 12} (2016) 053}, \href{https://arxiv.org/abs/1502.06120}{{\ttfamily arXiv:1502.06120 [hep-th]}}.

\bibitem{Nichols:2018qac}
D.~A. Nichols, ``{Center-of-mass angular momentum and memory effect in asymptotically flat spacetimes},'' \href{https://dx.doi.org/10.1103/PhysRevD.98.064032}{{\em Phys. Rev. D} {\bfseries 98} no.~6, (2018) 064032}, \href{https://arxiv.org/abs/1807.08767}{{\ttfamily arXiv:1807.08767 [gr-qc]}}.

\bibitem{DeLuca:2024bpt}
V.~De~Luca, J.~Khoury, and S.~S.~C. Wong, ``{Gravitational memory and soft theorems: the local perspective},'' \href{https://arxiv.org/abs/2412.01910}{{\ttfamily arXiv:2412.01910 [gr-qc]}}.

\bibitem{Strominger:2014pwa}
A.~Strominger and A.~Zhiboedov, ``{Gravitational Memory, BMS Supertranslations and Soft Theorems},'' \href{https://dx.doi.org/10.1007/JHEP01(2016)086}{{\em JHEP} {\bfseries 01} (2016) 086}, \href{https://arxiv.org/abs/1411.5745}{{\ttfamily arXiv:1411.5745 [hep-th]}}.

\bibitem{Strominger:2017zoo}
A.~Strominger, {\em {Lectures on the Infrared Structure of Gravity and Gauge Theory}}.
\newblock 3, 2017.

\bibitem{Wiseman:1991ss}
A.~G. Wiseman and C.~M. Will, ``{Christodoulou's nonlinear gravitational wave memory: Evaluation in the quadrupole approximation},'' \href{https://dx.doi.org/10.1103/PhysRevD.44.R2945}{{\em Phys. Rev. D} {\bfseries 44} no.~10, (1991) R2945--R2949}.

\bibitem{Favata:2010zu}
M.~Favata, ``{The gravitational-wave memory effect},'' \href{https://dx.doi.org/10.1088/0264-9381/27/8/084036}{{\em Class. Quantum Gravity} {\bfseries 27} (2010) 084036}, \href{https://arxiv.org/abs/1003.3486}{{\ttfamily arXiv:1003.3486 [gr-qc]}}.

\bibitem{Heisenberg:2023prj}
L.~Heisenberg, N.~Yunes, and J.~Zosso, ``{Gravitational wave memory beyond general relativity},'' \href{https://dx.doi.org/10.1103/PhysRevD.108.024010}{{\em Phys. Rev. D} {\bfseries 108} no.~2, (2023) 024010}.

\bibitem{PhysRev.121.1556}
R.~Arnowitt, S.~Deser, and C.~W. Misner, ``{Wave Zone in General Relativity},'' \href{https://dx.doi.org/10.1103/PhysRev.121.1556}{{\em Phys. Rev.} {\bfseries 121} (Mar, 1961) 1556--1566}.

\bibitem{PhysRev.166.1272}
R.~A. Isaacson, ``Gravitational radiation in the limit of high frequency. ii. nonlinear terms and the effective stress tensor,'' \href{https://dx.doi.org/10.1103/PhysRev.166.1272}{{\em Phys. Rev.} {\bfseries 166} (Feb, 1968) 1272--1280}. \url{https://link.aps.org/doi/10.1103/PhysRev.166.1272}.

\bibitem{Favata:2011qi}
M.~Favata, ``{The Gravitational-wave memory from eccentric binaries},'' \href{https://dx.doi.org/10.1103/PhysRevD.84.124013}{{\em Phys. Rev. D} {\bfseries 84} (2011) 124013}, \href{https://arxiv.org/abs/1108.3121}{{\ttfamily arXiv:1108.3121 [gr-qc]}}.

\bibitem{PhysRevD.98.064031}
C.~Talbot, E.~Thrane, P.~D. Lasky, and F.~Lin, ``Gravitational-wave memory: Waveforms and phenomenology,'' \href{https://dx.doi.org/10.1103/PhysRevD.98.064031}{{\em Phys. Rev. D} {\bfseries 98} (Sep, 2018) 064031}. \url{https://link.aps.org/doi/10.1103/PhysRevD.98.064031}.

\bibitem{PhysRevD.99.064045}
V.~Varma, S.~E. Field, M.~A. Scheel, J.~Blackman, L.~E. Kidder, and H.~P. Pfeiffer, ``Surrogate model of hybridized numerical relativity binary black hole waveforms,'' \href{https://dx.doi.org/10.1103/PhysRevD.99.064045}{{\em Phys. Rev. D} {\bfseries 99} (Mar, 2019) 064045}. \url{https://link.aps.org/doi/10.1103/PhysRevD.99.064045}.

\bibitem{Pollney_2011}
D.~Pollney and C.~Reisswig, ``Gravitational memory in binary black hole mergers,'' \href{https://dx.doi.org/10.1088/2041-8205/732/1/L13}{{\em The Astrophysical Journal Letters} {\bfseries 732} no.~1, (Apr, 2010) L13}. \url{https://dx.doi.org/10.1088/2041-8205/732/1/L13}.

\bibitem{Gasparotto:2023fcg}
S.~Gasparotto, R.~Vicente, D.~Blas, A.~C. Jenkins, and E.~Barausse, ``{Can gravitational-wave memory help constrain binary black-hole parameters? A LISA case study},'' \href{https://dx.doi.org/10.1103/PhysRevD.107.124033}{{\em Phys. Rev. D} {\bfseries 107} no.~12, (2023) 124033}, \href{https://arxiv.org/abs/2301.13228}{{\ttfamily arXiv:2301.13228 [gr-qc]}}.

\bibitem{Agazie:2025oug}
G.~Agazie {\em et~al.}, ``{The NANOGrav 15-year Data Set: Search for Gravitational Wave Memory},'' \href{https://arxiv.org/abs/2502.18599}{{\ttfamily arXiv:2502.18599 [gr-qc]}}.

\bibitem{NANOGrav:2019vto}
{\bfseries NANOGrav} Collaboration, K.~Aggarwal {\em et~al.}, ``{The NANOGrav 11 yr Data Set: Limits on Gravitational Wave Memory},'' \href{https://dx.doi.org/10.3847/1538-4357/ab6083}{{\em Astrophys. J.} {\bfseries 889} (2020) 38}, \href{https://arxiv.org/abs/1911.08488}{{\ttfamily arXiv:1911.08488 [astro-ph.HE]}}.

\bibitem{NANOGrav:2023vfo}
{\bfseries NANOGrav} Collaboration, G.~Agazie {\em et~al.}, ``{The NANOGrav 12.5 yr Data Set: Search for Gravitational Wave Memory},'' \href{https://dx.doi.org/10.3847/1538-4357/ad0726}{{\em Astrophys. J.} {\bfseries 963} no.~1, (2024) 61}, \href{https://arxiv.org/abs/2307.13797}{{\ttfamily arXiv:2307.13797 [gr-qc]}}.

\bibitem{Klimenko:2008fu}
S.~Klimenko, I.~Yakushin, A.~Mercer, and G.~Mitselmakher, ``{Coherent method for detection of gravitational wave bursts},'' \href{https://dx.doi.org/10.1088/0264-9381/25/11/114029}{{\em Class. Quant. Grav.} {\bfseries 25} (2008) 114029}, \href{https://arxiv.org/abs/0802.3232}{{\ttfamily arXiv:0802.3232 [gr-qc]}}.

\bibitem{Klimenko:2015ypf}
S.~Klimenko {\em et~al.}, ``{Method for detection and reconstruction of gravitational wave transients with networks of advanced detectors},'' \href{https://dx.doi.org/10.1103/PhysRevD.93.042004}{{\em Phys. Rev. D} {\bfseries 93} no.~4, (2016) 042004}, \href{https://arxiv.org/abs/1511.05999}{{\ttfamily arXiv:1511.05999 [gr-qc]}}.

\bibitem{Inchauspe:2024ibs}
H.~Inchausp\'e, S.~Gasparotto, D.~Blas, L.~Heisenberg, J.~Zosso, and S.~Tiwari, ``{Measuring gravitational wave memory with LISA},'' \href{https://arxiv.org/abs/2406.09228}{{\ttfamily arXiv:2406.09228 [gr-qc]}}.

\bibitem{Maggiore:1900zz}
M.~Maggiore, \href{https://dx.doi.org/10.1093/acprof:oso/9780198570745.001.0001}{{\em {Gravitational Waves. Vol. 1: Theory and Experiments}}}.
\newblock Oxford University Press, 2007.

\bibitem{Antelis:2018sfj}
J.~M. Antelis, J.~M. Hern\'andez, and C.~Moreno, ``{Post-Newtonian approximation of gravitational waves from the inspiral phase},'' \href{https://dx.doi.org/10.1088/1742-6596/1030/1/012005}{{\em J. Phys. Conf. Ser.} {\bfseries 1030} no.~1, (2018) 012005}.

\bibitem{Maggiore:2007ulw}
M.~Maggiore, {\em {Gravitational Waves. Vol. 1: Theory and Experiments}}.
\newblock Oxford Master Series in Physics. Oxford University Press, 2007.

\bibitem{Robson:2018ifk}
T.~Robson, N.~J. Cornish, and C.~Liu, ``{The construction and use of LISA sensitivity curves},'' \href{https://dx.doi.org/10.1088/1361-6382/ab1101}{{\em Class. Quant. Grav.} {\bfseries 36} no.~10, (2019) 105011}, \href{https://arxiv.org/abs/1803.01944}{{\ttfamily arXiv:1803.01944 [astro-ph.HE]}}.

\bibitem{1931ApJ....74...81C}
S.~Chandrasekhar, ``{The maximum mass of ideal white dwarfs},'' \href{https://dx.doi.org/10.1086/143324}{{\em Astrophys. J.} {\bfseries 74} (1931) 81--82}.

\bibitem{Metzger:2024ujc}
B.~D. Metzger, L.~Hui, and M.~Cantiello, ``{Fragmentation in Gravitationally Unstable Collapsar Disks and Subsolar Neutron Star Mergers},'' \href{https://dx.doi.org/10.3847/2041-8213/ad6990}{{\em Astrophys. J. Lett.} {\bfseries 971} no.~2, (2024) L34}, \href{https://arxiv.org/abs/2407.07955}{{\ttfamily arXiv:2407.07955 [astro-ph.HE]}}.

\bibitem{Muller:2024aod}
B.~M\"uller, A.~Heger, and J.~Powell, ``{Minimum Neutron Star Mass in Neutrino-Driven Supernova Explosions},'' \href{https://dx.doi.org/10.1103/PhysRevLett.134.071403}{{\em Phys. Rev. Lett.} {\bfseries 134} no.~7, (2025) 071403}, \href{https://arxiv.org/abs/2407.08407}{{\ttfamily arXiv:2407.08407 [astro-ph.HE]}}.

\bibitem{Cardoso:2019rvt}
V.~Cardoso and P.~Pani, ``{Testing the nature of dark compact objects: a status report},'' \href{https://dx.doi.org/10.1007/s41114-019-0020-4}{{\em Living Rev. Rel.} {\bfseries 22} no.~1, (2019) 4}, \href{https://arxiv.org/abs/1904.05363}{{\ttfamily arXiv:1904.05363 [gr-qc]}}.

\bibitem{Coleman:1985ki}
S.~R. Coleman, ``{Q-balls},'' \href{https://dx.doi.org/10.1016/0550-3213(86)90520-1}{{\em Nucl. Phys. B} {\bfseries 262} no.~2, (1985) 263}. [Addendum: Nucl.Phys.B 269, 744 (1986)].

\bibitem{Colpi:1986ye}
M.~Colpi, S.~L. Shapiro, and I.~Wasserman, ``{Boson Stars: Gravitational Equilibria of Selfinteracting Scalar Fields},'' \href{https://dx.doi.org/10.1103/PhysRevLett.57.2485}{{\em Phys. Rev. Lett.} {\bfseries 57} (1986) 2485--2488}.

\bibitem{Liebling:2012fv}
S.~L. Liebling and C.~Palenzuela, ``{Dynamical boson stars},'' \href{https://dx.doi.org/10.1007/s41114-023-00043-4}{{\em Living Rev. Rel.} {\bfseries 26} no.~1, (2023) 1}, \href{https://arxiv.org/abs/1202.5809}{{\ttfamily arXiv:1202.5809 [gr-qc]}}.

\bibitem{Lee:1986tr}
T.~D. Lee and Y.~Pang, ``{Fermion Soliton Stars and Black Holes},'' \href{https://dx.doi.org/10.1103/PhysRevD.35.3678}{{\em Phys. Rev. D} {\bfseries 35} (1987) 3678}.

\bibitem{DelGrosso:2023trq}
L.~Del~Grosso, G.~Franciolini, P.~Pani, and A.~Urbano, ``{Fermion soliton stars},'' \href{https://dx.doi.org/10.1103/PhysRevD.108.044024}{{\em Phys. Rev. D} {\bfseries 108} no.~4, (2023) 044024}, \href{https://arxiv.org/abs/2301.08709}{{\ttfamily arXiv:2301.08709 [gr-qc]}}.

\bibitem{DelGrosso:2023dmv}
L.~Del~Grosso and P.~Pani, ``{Fermion soliton stars with asymmetric vacua},'' \href{https://dx.doi.org/10.1103/PhysRevD.108.064042}{{\em Phys. Rev. D} {\bfseries 108} no.~6, (2023) 064042}, \href{https://arxiv.org/abs/2308.15921}{{\ttfamily arXiv:2308.15921 [gr-qc]}}.

\bibitem{Raidal:2024bmm}
M.~Raidal, V.~Vaskonen, and H.~Veerm\"ae, {\em {Formation of primordial black hole binaries and their merger rates}}.
\newblock 4, 2024.
\newblock \href{https://arxiv.org/abs/2404.08416}{{\ttfamily arXiv:2404.08416 [astro-ph.CO]}}.

\bibitem{Hutsi:2020sol}
G.~H\"utsi, M.~Raidal, V.~Vaskonen, and H.~Veerm\"ae, ``{Two populations of LIGO-Virgo black holes},'' \href{https://dx.doi.org/10.1088/1475-7516/2021/03/068}{{\em JCAP} {\bfseries 03} (2021) 068}, \href{https://arxiv.org/abs/2012.02786}{{\ttfamily arXiv:2012.02786 [astro-ph.CO]}}.

\bibitem{Liu:2018ess}
L.~Liu, Z.-K. Guo, and R.-G. Cai, ``{Effects of the surrounding primordial black holes on the merger rate of primordial black hole binaries},'' \href{https://dx.doi.org/10.1103/PhysRevD.99.063523}{{\em Phys. Rev. D} {\bfseries 99} no.~6, (2019) 063523}, \href{https://arxiv.org/abs/1812.05376}{{\ttfamily arXiv:1812.05376 [astro-ph.CO]}}.

\bibitem{Crescimbeni:2025ywm}
F.~Crescimbeni, V.~Desjacques, G.~Franciolini, A.~Ianniccari, A.~J. Iovino, G.~Perna, D.~Perrone, A.~Riotto, and H.~Veerm\"ae, ``{The Irrelevance of Primordial Black Hole Clustering in the LVK mass range},'' \href{https://arxiv.org/abs/2502.01617}{{\ttfamily arXiv:2502.01617 [astro-ph.CO]}}.

\bibitem{Inman:2019wvr}
D.~Inman and Y.~Ali-Ha\"\i{}moud, ``{Early structure formation in primordial black hole cosmologies},'' \href{https://dx.doi.org/10.1103/PhysRevD.100.083528}{{\em Phys. Rev. D} {\bfseries 100} no.~8, (2019) 083528}, \href{https://arxiv.org/abs/1907.08129}{{\ttfamily arXiv:1907.08129 [astro-ph.CO]}}.

\bibitem{DeLuca:2020jug}
V.~De~Luca, V.~Desjacques, G.~Franciolini, and A.~Riotto, ``{The clustering evolution of primordial black holes},'' \href{https://dx.doi.org/10.1088/1475-7516/2020/11/028}{{\em JCAP} {\bfseries 11} (2020) 028}, \href{https://arxiv.org/abs/2009.04731}{{\ttfamily arXiv:2009.04731 [astro-ph.CO]}}.

\bibitem{Pujolas:2021yaw}
O.~Pujolas, V.~Vaskonen, and H.~Veerm\"ae, ``{Prospects for probing gravitational waves from primordial black hole binaries},'' \href{https://dx.doi.org/10.1103/PhysRevD.104.083521}{{\em Phys. Rev. D} {\bfseries 104} no.~8, (2021) 083521}, \href{https://arxiv.org/abs/2107.03379}{{\ttfamily arXiv:2107.03379 [astro-ph.CO]}}.

\bibitem{1989ApJ...343..725Q}
G.~D. {Quinlan} and S.~L. {Shapiro}, ``{Dynamical Evolution of Dense Clusters of Compact Stars},'' \href{https://dx.doi.org/10.1086/167745}{{\em \apj} {\bfseries 343} (Aug., 1989) 725}.

\bibitem{Mouri:2002mc}
H.~Mouri and Y.~Taniguchi, ``{Runaway merging of black holes: analytical constraint on the timescale},'' \href{https://dx.doi.org/10.1086/339472}{{\em Astrophys. J. Lett.} {\bfseries 566} (2002) L17--L20}, \href{https://arxiv.org/abs/astro-ph/0201102}{{\ttfamily arXiv:astro-ph/0201102}}.

\bibitem{OLeary:2008myb}
R.~M. O'Leary, B.~Kocsis, and A.~Loeb, ``{Gravitational waves from scattering of stellar-mass black holes in galactic nuclei},'' \href{https://dx.doi.org/10.1111/j.1365-2966.2009.14653.x}{{\em Mon. Not. Roy. Astron. Soc.} {\bfseries 395} no.~4, (2009) 2127--2146}, \href{https://arxiv.org/abs/0807.2638}{{\ttfamily arXiv:0807.2638 [astro-ph]}}.

\bibitem{Korol:2019jud}
V.~Korol, I.~Mandel, M.~C. Miller, R.~P. Church, and M.~B. Davies, ``{Merger rates in primordial black hole clusters without initial binaries},'' \href{https://dx.doi.org/10.1093/mnras/staa1644}{{\em Mon. Not. Roy. Astron. Soc.} {\bfseries 496} no.~1, (2020) 994--1000}, \href{https://arxiv.org/abs/1911.03483}{{\ttfamily arXiv:1911.03483 [astro-ph.HE]}}.

\bibitem{Kritos:2020wcl}
K.~Kritos, V.~De~Luca, G.~Franciolini, A.~Kehagias, and A.~Riotto, ``{The Astro-Primordial Black Hole Merger Rates: a Reappraisal},'' \href{https://dx.doi.org/10.1088/1475-7516/2021/05/039}{{\em JCAP} {\bfseries 05} (2021) 039}, \href{https://arxiv.org/abs/2012.03585}{{\ttfamily arXiv:2012.03585 [gr-qc]}}.

\bibitem{Franciolini:2022ewd}
G.~Franciolini, K.~Kritos, E.~Berti, and J.~Silk, ``{Primordial black hole mergers from three-body interactions},'' \href{https://dx.doi.org/10.1103/PhysRevD.106.083529}{{\em Phys. Rev. D} {\bfseries 106} no.~8, (2022) 083529}, \href{https://arxiv.org/abs/2205.15340}{{\ttfamily arXiv:2205.15340 [astro-ph.CO]}}.

\bibitem{Franciolini:2021xbq}
G.~Franciolini, R.~Cotesta, N.~Loutrel, E.~Berti, P.~Pani, and A.~Riotto, ``{How to assess the primordial origin of single gravitational-wave events with mass, spin, eccentricity, and deformability measurements},'' \href{https://dx.doi.org/10.1103/PhysRevD.105.063510}{{\em Phys. Rev. D} {\bfseries 105} no.~6, (2022) 063510}, \href{https://arxiv.org/abs/2112.10660}{{\ttfamily arXiv:2112.10660 [astro-ph.CO]}}.

\bibitem{Planck:2015fie}
{\bfseries Planck} Collaboration, P.~A.~R. Ade {\em et~al.}, ``{Planck 2015 results. XIII. Cosmological parameters},'' \href{https://dx.doi.org/10.1051/0004-6361/201525830}{{\em Astron. Astrophys.} {\bfseries 594} (2016) A13}, \href{https://arxiv.org/abs/1502.01589}{{\ttfamily arXiv:1502.01589 [astro-ph.CO]}}.

\bibitem{KAGRA:2013rdx}
{\bfseries KAGRA, LIGO Scientific, Virgo} Collaboration, B.~P. Abbott {\em et~al.}, ``{Prospects for observing and localizing gravitational-wave transients with Advanced LIGO, Advanced Virgo and KAGRA},'' \href{https://dx.doi.org/10.1007/s41114-020-00026-9}{{\em Living Rev. Rel.} {\bfseries 19} (2016) 1}, \href{https://arxiv.org/abs/1304.0670}{{\ttfamily arXiv:1304.0670 [gr-qc]}}.

\bibitem{Hild:2008ng}
S.~Hild, S.~Chelkowski, and A.~Freise, ``{Pushing towards the ET sensitivity using 'conventional' technology},'' \href{https://arxiv.org/abs/0810.0604}{{\ttfamily arXiv:0810.0604 [gr-qc]}}.

\bibitem{Punturo:2010zz}
M.~Punturo {\em et~al.}, ``{The Einstein Telescope: A third-generation gravitational wave observatory},'' \href{https://dx.doi.org/10.1088/0264-9381/27/19/194002}{{\em Class. Quant. Grav.} {\bfseries 27} (2010) 194002}.

\bibitem{Hild:2010id}
S.~Hild {\em et~al.}, ``{Sensitivity Studies for Third-Generation Gravitational Wave Observatories},'' \href{https://dx.doi.org/10.1088/0264-9381/28/9/094013}{{\em Class. Quant. Grav.} {\bfseries 28} (2011) 094013}, \href{https://arxiv.org/abs/1012.0908}{{\ttfamily arXiv:1012.0908 [gr-qc]}}.

\bibitem{LISA:2017pwj}
{\bfseries LISA} Collaboration, P.~Amaro-Seoane {\em et~al.}, ``{Laser Interferometer Space Antenna},'' \href{https://arxiv.org/abs/1702.00786}{{\ttfamily arXiv:1702.00786 [astro-ph.IM]}}.

\bibitem{LISACosmologyWorkingGroup:2022jok}
{\bfseries LISA Cosmology Working Group} Collaboration, P.~Auclair {\em et~al.}, ``{Cosmology with the Laser Interferometer Space Antenna},'' \href{https://dx.doi.org/10.1007/s41114-023-00045-2}{{\em Living Rev. Rel.} {\bfseries 26} no.~1, (2023) 5}, \href{https://arxiv.org/abs/2204.05434}{{\ttfamily arXiv:2204.05434 [astro-ph.CO]}}.

\bibitem{Domcke:2024mfu}
V.~Domcke, S.~A.~R. Ellis, and N.~L. Rodd, ``{Magnets are Weber Bar Gravitational Wave Detectors},'' \href{https://arxiv.org/abs/2408.01483}{{\ttfamily arXiv:2408.01483 [hep-ph]}}.

\bibitem{DeLuca:2021hde}
V.~De~Luca, G.~Franciolini, P.~Pani, and A.~Riotto, ``{The minimum testable abundance of primordial black holes at future gravitational-wave detectors},'' \href{https://dx.doi.org/10.1088/1475-7516/2021/11/039}{{\em JCAP} {\bfseries 11} (2021) 039}, \href{https://arxiv.org/abs/2106.13769}{{\ttfamily arXiv:2106.13769 [astro-ph.CO]}}.

\bibitem{Miller:2024khl}
A.~L. Miller, ``{Prospects for detecting asteroid-mass primordial black holes in extreme-mass-ratio inspirals with continuous gravitational waves},'' \href{https://arxiv.org/abs/2410.01348}{{\ttfamily arXiv:2410.01348 [gr-qc]}}.

\bibitem{Allen:2019hnd}
B.~Allen, ``{Gravitational wave stochastic background from cosmological particle decay},'' \href{https://dx.doi.org/10.1103/PhysRevResearch.2.012034}{{\em Phys. Rev. Res.} {\bfseries 2} no.~1, (2020) 012034}, \href{https://arxiv.org/abs/1910.08213}{{\ttfamily arXiv:1910.08213 [gr-qc]}}.

\bibitem{Zhao:2021zlr}
Z.-C. Zhao and Z.~Cao, ``{Stochastic gravitational wave background due to gravitational wave memory},'' \href{https://dx.doi.org/10.1007/s11433-022-1965-y}{{\em Sci. China Phys. Mech. Astron.} {\bfseries 65} no.~11, (2022) 119511}, \href{https://arxiv.org/abs/2111.13883}{{\ttfamily arXiv:2111.13883 [gr-qc]}}.

\bibitem{Boybeyi:2024aax}
T.~Boybeyi, V.~Mandic, and A.~Papageorgiou, ``{Astrometric deflections from gravitational wave memory accumulation over cosmological scales},'' \href{https://dx.doi.org/10.1103/PhysRevD.110.043047}{{\em Phys. Rev. D} {\bfseries 110} no.~4, (2024) 043047}, \href{https://arxiv.org/abs/2403.07614}{{\ttfamily arXiv:2403.07614 [astro-ph.CO]}}.

\bibitem{1977ApJ...216..610T}
M.~{Turner}, ``{Gravitational radiation from point-masses in unbound orbits: Newtonian results.},'' \href{https://dx.doi.org/10.1086/155501}{{\em \apj} {\bfseries 216} (Sept., 1977) 610--619}.

\bibitem{Braginsky:1987kwo}
V.~B. Braginsky and K.~S. Thorne, ``{Gravitational-wave bursts with memory and experimental prospects},'' \href{https://dx.doi.org/10.1038/327123a0}{{\em Nature} {\bfseries 327} (1987) 123--125}.

\bibitem{Caldarola:2023ipo}
M.~Caldarola, S.~Kuroyanagi, S.~Nesseris, and J.~Garcia-Bellido, ``{Effects of orbital precession on hyperbolic encounters},'' \href{https://dx.doi.org/10.1103/PhysRevD.109.064001}{{\em Phys. Rev. D} {\bfseries 109} no.~6, (2024) 064001}, \href{https://arxiv.org/abs/2307.00915}{{\ttfamily arXiv:2307.00915 [gr-qc]}}.

\bibitem{Damour:2000wa}
T.~Damour and A.~Vilenkin, ``{Gravitational wave bursts from cosmic strings},'' \href{https://dx.doi.org/10.1103/PhysRevLett.85.3761}{{\em Phys. Rev. Lett.} {\bfseries 85} (2000) 3761--3764}, \href{https://arxiv.org/abs/gr-qc/0004075}{{\ttfamily arXiv:gr-qc/0004075}}.

\bibitem{Amaro-Seoane:2007osp}
P.~Amaro-Seoane, J.~R. Gair, M.~Freitag, M.~Coleman~Miller, I.~Mandel, C.~J. Cutler, and S.~Babak, ``{Astrophysics, detection and science applications of intermediate- and extreme mass-ratio inspirals},'' \href{https://dx.doi.org/10.1088/0264-9381/24/17/R01}{{\em Class. Quant. Grav.} {\bfseries 24} (2007) R113--R169}, \href{https://arxiv.org/abs/astro-ph/0703495}{{\ttfamily arXiv:astro-ph/0703495}}.

\bibitem{Favata:2008yd}
M.~Favata, ``{Post-Newtonian corrections to the gravitational-wave memory for quasi-circular, inspiralling compact binaries},'' \href{https://dx.doi.org/10.1103/PhysRevD.80.024002}{{\em Phys. Rev. D} {\bfseries 80} (2009) 024002}, \href{https://arxiv.org/abs/0812.0069}{{\ttfamily arXiv:0812.0069 [gr-qc]}}.

\bibitem{Zosso:2024xgy}
J.~Zosso, \href{https://dx.doi.org/10.3929/ethz-b-000675938}{{\em {Probing Gravity - Fundamental Aspects of Metric Theories and their Implications for Tests of General Relativity}}}.
\newblock PhD thesis, Zurich, ETH, 2024.
\newblock \href{https://arxiv.org/abs/2412.06043}{{\ttfamily arXiv:2412.06043 [gr-qc]}}.

\bibitem{favata_nonlinear_2009}
M.~Favata, ``Nonlinear gravitational-wave memory from binary black hole mergers,'' \href{https://dx.doi.org/10.1088/0004-637X/696/2/L159}{{\em The Astrophysical Journal} {\bfseries 696} no.~2, (Apr., 2009) L159--L162}. \url{https://doi.org/10.1088/0004-637x/696/2/l159}. Publisher: American Astronomical Society.

\bibitem{Romano:2016dpx}
J.~D. Romano and N.~J. Cornish, ``{Detection methods for stochastic gravitational-wave backgrounds: a unified treatment},'' \href{https://dx.doi.org/10.1007/s41114-017-0004-1}{{\em Living Rev. Rel.} {\bfseries 20} no.~1, (2017) 2}, \href{https://arxiv.org/abs/1608.06889}{{\ttfamily arXiv:1608.06889 [gr-qc]}}.

\bibitem{Barsotti:18}
L.~Barsotti, S.~Gras, M.~Evans, and P.~Fritschel, {\em The updated Advanced LIGO design curve}.
\newblock LIGO, 2018.
\newblock \url{https://dcc.ligo.org/LIGO-T1800044/public}.
\newblock {L}IGO Document T1800044.

\bibitem{DMRadio:2022jfv}
{\bfseries DMRadio} Collaboration, L.~Brouwer {\em et~al.}, ``{Proposal for a definitive search for GUT-scale QCD axions},'' \href{https://dx.doi.org/10.1103/PhysRevD.106.112003}{{\em Phys. Rev. D} {\bfseries 106} no.~11, (2022) 112003}, \href{https://arxiv.org/abs/2203.11246}{{\ttfamily arXiv:2203.11246 [hep-ex]}}.

\end{thebibliography}\endgroup
\end{document}